\newcommand{\arxiv}[1]{#1}
\newcommand{\oops}[1]{}

\arxiv{
\documentclass[12pt]{article}
\usepackage{fullpage}
}

\oops{}

\usepackage{comment}
\newif\ifoopsenv
\oopsenvfalse  %
\ifoopsenv
  \newenvironment{oopsenv}{}{}
\else
  \excludecomment{oopsenv}
\fi

\newif\ifarxivenv
\arxivenvtrue    %
\ifarxivenv
  \newenvironment{arxivenv}{}{}
\else
  \excludecomment{arxivenv}
\fi

\usepackage{float}

\usepackage{hyperref} %
\usepackage{xspace}
\usepackage{alltt}

\usepackage{enumitem}
\setlist[description]{itemindent=-2ex}
\oops{}

\usepackage[dvipsnames]{xcolor}
\definecolor{codegray}{rgb}{0.5,0.5,0.5}
\definecolor{codepurple}{rgb}{0.58,0,0.82}

\arxiv{\usepackage{amsmath,amssymb}}
\usepackage{graphicx}
\usepackage{multirow}
\usepackage{multicol}
\usepackage{setspace} %

\usepackage{tikz}
\usepackage{pgfplots}
\usepackage{csvsimple}
\usepackage{filecontents}
\usepackage{pgfplotstable} %
\pgfplotsset{compat=1.15}

\usepackage{listings}

\lstdefinestyle{mystyle}{
    backgroundcolor=\color{white},   
    basicstyle=\arxiv{\footnotesize}\oops{}\ttfamily,
    commentstyle=\color{brown},
    keywordstyle=\color{blue}, %
    stringstyle=\color{codepurple},
    numberstyle=\tiny\color{codegray},
    numbers=none,
    breakatwhitespace=false,
    breaklines=false,                 
    captionpos=b,
    keepspaces=true,                 
    numbersep=5pt,
    showspaces=false,                
    showstringspaces=false,
    showtabs=false,                  
    tabsize=2
}
\newcommand{\Hex}[1]{\hspace{#1ex}}
\newcommand{\Vex}[1]{\vspace{#1ex}}

\newenvironment{code}{\Vex{.5}\begin{alltt}\arxiv{\small}\oops{}}{\end{alltt}\Vex{.5}}
\newcommand\co[1]{\mbox{\small\tt #1}} %
\newcommand\kw[1]{\textcolor{blue}{#1}} %
\newcommand{\mathify}[1]{\ifmmode{\mbox{$#1$}}\else\mbox{$#1$}\fi}
\newcommand\m[1]{$#1$} %
\newcommand\p[1]{\mathify{\it #1}} %
\newcommand\bigO[1]{\m{O(#1)}} %

\newcommand\WHILE{\kw{while}\xspace}
\newcommand\INFER{\kw{infer}}
\newcommand\RULES{\kw{rules}}
\newcommand\cIF{\kw{if}\xspace}

\newcommand\NOT{\kw{not}\xspace}
\newcommand\SOME{\kw{some}\xspace}

\newcommand\IN{\kw{in}\xspace}

\newcommand{\mysubsec}[1]{\subsection{#1}}
\newcommand{\mypar}[1]{\Vex{2} \noindent {\bf #1.~}}

\newenvironment{example}{\Vex{.5}\par\textbf{\textit{Example}}.}{\hfill$\blacksquare$\par}

\newcommand\defn[1]{{\it #1}}

\newcommand\myurl[1]{{\footnotesize\url{#1}}}
\newcommand\ghurl[2]{} %
\newcommand\ver[1]{\co{#1}} %

\newcommand{\notes}[1]{}

\usepackage{longtable}

\date{May 29, 2022}

\newcommand\fund{
    This work was supported in part by NSF under grants 
    CCF-1954837, %
    CCF-1414078, and %
    IIS-1447549 %
    and ONR under grants
    N00014-21-1-2719, %
    N00014-20-1-2751, and %
    N00014-15-1-2208. %
}

\begin{document}

\title{Programming with rules and everything else, seamlessly%
\thanks{\fund}} 
\author%
{Yanhong A.\ Liu \Hex{1.5} Scott D.\ Stoller \Hex{1.5} 
    Yi Tong \Hex{1.5} Bo Lin \Hex{1.5} K.\,Tuncay Tekle\Vex{1}\\
  Stony Brook University\\
  \{liu,stoller,yittong,bolin,tuncay\}@cs.stonybrook.edu}

\oops{}

\arxiv{
\maketitle
}
\Vex{-4}

\begin{abstract}

Logic rules are powerful for expressing complex reasoning and analysis
problems.  At the same time, they are inconvenient or impossible to use
for many other aspects of applications.  Integrating rules in a language
with sets and functions, and furthermore with updates to objects, has been
a subject of significant study.  What's lacking is a language that
integrates all constructs seamlessly.

This paper presents a language, Alda, that supports all of rules,
sets, functions, updates, and objects as seamlessly integrated built-ins, including concurrent and distributed processes.
The key idea is to support predicates as set-valued variables that can be used and updated in any scope, and support queries and inference with both explicit and automatic calls to an inference function.  We develop a complete formal semantics for Alda.%
\notes{}
We design a compilation framework that ensures the declarative
semantics of rules, while also being able to exploit available optimizations.
We describe a prototype implementation that builds on a powerful extension
of Python and employs an efficient logic rule engine.
We develop a range of benchmarks and present results of experiments
to demonstrate Alda's power for programming and generally good performance.

\end{abstract}

\oops{}

\section{Introduction}

Logic rules and inference are powerful for specifying and solving complex
reasoning and analysis problems~\cite{liu18LPappl-wbook}, especially 
in critical areas such as program analysis, %
decision support, networking, and security~\cite{WarLiu17AppLP-arxiv}.
Many logic rule languages and systems have been developed, and used
successfully in many areas of applications, expressing challenging problems
succinctly and solving them with general and powerful %
implementation methods~\cite{KifLiu18wbook}.

At the same time, logic rules are inconvenient or impossible to use for
many other aspects of applications---computations that are better specified using set queries, recursive functions, state updates including input/output
operations, and object encapsulation.

Significant effort has been devoted to integrating or interfacing logic
programming with
other programming paradigms---database programming, functional programming,
imperative programming, and object-oriented programming---%
resulting in many mixtures of languages~\cite{KifLiu18wbook,prolog50tplp}.
What's challenging has been:
\begin{enumerate}
\item[1)] 
  a simple and powerful language that can express computations using all of
  rules, sets, functions, updates, and objects, %
  naturally as built-ins
  without extra interfaces, and with a clear integrated semantics; and
\item[2)]
  a compilation and optimization framework for this powerful language, 
  for implementation in a widely-used programming language, and with 
  generally good performance.
\end{enumerate}

This paper presents %
a powerful language, Alda, that allows computations to be expressed easily
and clearly using all of rules, sets, functions, updates, and objects as
well as concurrent and distributed processes.
\begin{itemize}

\item Sets of rules can be specified in any scope, 
  just as other built-in constructs in high-level object-oriented
  languages.

\item Predicates in rules are treated as set variables holding the set of 
  tuples for which the predicate is true, and vice versa.
  Thus, predicates can be used directly in other constructs as 
  variables without needing %
  any interface. %
  In this way, predicates are completely different from functions
  or procedures, unlike in previous logic languages and extensions.

\item Predicates as variables can be global variables, object fields, or
  local variables in a rule set.  They can be aliased if needed for
  efficiency, just like other variables.  Predicates as variables also
  avoid the need for higher-order predicates or more complicated features for reusing predicate definitions in
  more complex logic languages.

\item A dedicated inference function can be called any time on a rule set,
  to infer values of derived predicates (i.e., predicates in conclusions of
  rules) and answer queries, given values of base predicates (i.e., predicates
  not in conclusions of rules).

\item Declarative semantics of rules are maintained automatically after any
  update that may affect the meaning of the rules, including in the
  presence of aliasing, through implicit calls to the inference function
  as needed.
\end{itemize}
We also define a formal semantics that integrates declarative and
operational semantics.  The integrated semantics supports,
seamlessly, all of logic programming with rules, database programming
with sets, functional programming, imperative programming, and
object-oriented programming including concurrent and distributed
programming. %

Implementing such a powerful language to support this
diverse range of features is also challenging, especially to make all
features easily usable in a widely-used language.  Furthermore, the
conceptually simple semantics for ensuring declarative semantics requires
different implementation
methods depending on the kinds of updates in the program.

We design a compilation and optimization framework 
that ensures the
correct declarative semantics of rules, while also being able to exploit
available optimizations.%
\begin{itemize}
\item The compilation framework considers different kinds of updates and
  aliasing and uses the most efficient method for each kind.

\item The framework employs well-known optimizations to address different
  sources of potential inefficiencies.  This includes reusing results of
  expensive queries that use rules, updating derived predicates
  incrementally when base predicates are updated, as well as optimizations
  within the inference using rules. %
\end{itemize}

There has been a significant amount of related research, as discussed in
Section~\ref{sec-related}.  They either do not integrate rules with all of
sets, functions, updates, and objects, 
or require extra programming
interfaces to wrap features in special objects, pass code as special string
values, or convert data to special representations.

We also describe a prototype implementation of the language, programming and performance benchmarks, and experimental evaluation results.
These benchmarks and results help confirm the power and benefits of a
seamlessly integrated language and its generally good performance.

\arxiv{
The rest of the paper is organized as follows.
Section~\ref{sec-prob} describes the challenges of programming using rules with other features.
Section~\ref{sec-lang} presents the language and gives an overview of the
formal semantics.
Section~\ref{sec-impl} describes the compilation and optimization framework.
Section~\ref{sec-bench} presents programming and performance benchmarks.
Section~\ref{sec-expe} describes the implementation and experimental results.
Section~\ref{sec-related} discusses related work and concludes.
\oops{}\arxiv{Appendix~\ref{app-formal}}
presents a complete formal semantics for our language.
}

\section{Programming with rules and other features}
\label{sec-prob}

Logic rules and queries allow complex reasoning and analysis problems to be
expressed declaratively, easily and clearly, at a high level, often in ways
not possible using set expressions, recursive functions, or other
constructs.  However, logic rules are not appropriate for many other
aspects of applications, causing logic languages to include many
non-declarative features, including tricky special features
such as cut and negation as failure instead of logical
negation~\cite{Sterling:Shapiro:94}.

At the same time, languages commonly used in practice are imperative and
often object-oriented, because updates are essential in modeling the real
world,
and object encapsulation is fundamental for organizing
system components in large applications.  In these languages, logic rules
are missing, and tedious interfaces are needed to use them,
similar to how database interfaces such as
ODBC~\cite{geiger1995inside}
and JDBC~\cite{reese2000database}
are needed to use SQL queries.

The challenge is how to support rules with all other features seamlessly,
with no extra interfaces or boilerplate code.

\mypar{Running example}
We use the well-known transitive closure 
problem of graphs as a running example.

Given a predicate, \co{edge}, that asserts whether there is an edge from a
first vertex to a second vertex, the transitive closure problem defines a
predicate, \co{path},
that asserts whether there is a path 
from a first vertex 
to a second vertex by following the edges.
This can be expressed in dominant logic languages such as Prolog as follows:
\begin{lstlisting}
  path(X,Y) :- edge(X,Y).
  path(X,Y) :- edge(X,Z), path(Z,Y).
\end{lstlisting}
The first rule says that 
there is a path from \co{X} to \co{Y} if there is an edge from
\co{X} to \co{Y}. The second rule says that 
there is a path from \co{X} to \co{Y} if there is an edge from \co{X} to
\co{Z} and 
there is a path from \co{Z} to \co{Y}.  Then, one can query, for example,
\begin{itemize}
  \setlength{\itemsep}{0ex}
\item[1)] the transitive closure, that is, pairs of vertices where there is
  a path from the first vertex to the second vertex, by using
  \co{path(X,Y)},
\item[2)] vertices that are reachable from a given vertex, say vertex
  \co{1}, by \co{path(1,X)},
\item[3)] vertices that can reach a given vertex, say vertex \co{2}, by
  \co{path(X,2)}, and
\item[4)] whether a given vertex, say \co{1}, can reach a given vertex, say
  \co{2}, by \co{path(1,2)}.
\end{itemize}

\mypar{Advantages of using rules over everything else for complex queries}
It is easy to see that rules are declarative and powerful, making a complex
problem easy to express, with predicates (\co{edge}, \co{path}), logic
variables (\co{X}, \co{Y}, \co{Z}), constants (\co{1}, \co{2}), and a few
symbols.  In addition, the same rules can be used easily for different
kinds of queries.

Generalizations are easy too: a predicate may have more arguments, such as
a third argument for the weight of edges; there may be more kinds of edges,
expressing more kinds of relationships; and there can be more conditions,
called hypotheses, in a rule, as well as more rules.

By using efficient inference algorithms and implementation
methods~\cite{LiuSto03Rules-PPDP,LiuSto09Rules-TOPLAS}, specialization with
recursion conversion~\cite{Tek+08RulePE-AMAST}, and demand
transformation~\cite{TekLiu10RuleQuery-PPDP,TekLiu11RuleQueryBeat-SIGMOD},
queries using rules can have optimal complexities.
For example, given \m{m} edges over \m{n} vertices, the transitive closure
query can be \bigO{mn} time,
and the other three kinds of queries can be \bigO{m} time. 
With well-known logic rule engines, such as
XSB~\cite{SagSW94xsb,swift2012xsb,xsb21} with its efficient emulator in C,
queries using rules can be highly efficient, even close to 
manually written C programs.

Note that rules for solving a problem may be written in different ways.
For example, for the transitive closure problem, one may reverse the order
of the rules and, in the second rule, the order of the two hypotheses or
just the two predicate names, and one may change \co{edge} to \co{path} in
the second rule.  Even though existing highly optimized rule engines may
run with drastically different complexities for these different rules, the
optimizations described
above~\cite{LiuSto03Rules-PPDP,LiuSto09Rules-TOPLAS,Tek+08RulePE-AMAST,TekLiu10RuleQuery-PPDP,TekLiu11RuleQueryBeat-SIGMOD}
can give optimal complexities regardless of these different ways.

Without using rules, problems such as transitive closure %
could be solved by programming using imperative updates, set expressions,
recursive functions, and/or their combinations.  However, these programs are
drastically more complex or exceedingly inefficient or both.
\begin{itemize}

\item 
Using imperative updates with appropriate data structures, transitive
closure can be computed following a well-known \bigO{n^3} time algorithm,
or a better \bigO{mn} time algorithm~\cite{LiuSto09Rules-TOPLAS}.

However, it requires using an adjacency list or adjacency matrix for graph
representation, and a depth-first search or breadth-first search for
searching the graph, updating the detailed data structures carefully as
the search progresses.  The resulting program is orders of magnitude larger
and drastically more complex.

\item
Using sets and set expressions, an imperative algorithm can be written much
more simply at a higher level.  For example, in Python, given a set \co{E}
of edges, the transitive closure \co{T} can be computed as follows, where
\co{T} starts with \co{E} (using a copy so \co{E} is not changed when
\co{T} is), workset \co{W} keeps newly discovered pairs, and \co{\WHILE
  W} continues if \co{W} is not empty:
\begin{lstlisting}
  T = E.copy()
  W = {(x,y) for (x,z) in T for (z2,y) in E if z2==z} - T
  while W:
    T.add(W.pop())
    W = {(x,y) for (x,z) in T for (z2,y) in E if z2==z} - T
\end{lstlisting}
However, expensive set operations are computed for \co{W} in each
iteration, and the total is worst-case \bigO{mn^4} time.  One may find ways
to avoid the duplicated code for computing \co{W}, but similar set
operations in each iteration cannot be avoided.
Incrementalization~\cite{PaiKoe82,Liu13book} can derive efficient \bigO{mn}
time algorithm, but such transformation is not supported in general in
any commonly-used language.

\item
Using functions, one can wrap the imperative code above, either using
high-level sets or not, in a function definition that can be called at
uses.
\notes{}

However, if imperative updates and \co{\WHILE} loops are not allowed, the
resulting programs that use recursion would not be fundamentally easier or
simpler than the programs above.  In fact, writing these functional
programs are so nontrivial that it required research papers for individual
problems, e.g., for computing the transitive
closure~\cite{berghammer2015combining}, which develops a Haskell program
with no better than \bigO{m^3} time, which is worst-case \bigO{n^6}.

\item 
Using recursive queries in SQL, which are increasingly supported and are
essentially set expressions plus recursion, the transitive closure query
can be expressed more declaratively than the programs above.

However, using recursive SQL queries is much more complex than using
rules~\cite{KifBL06}.
Additionally, tedious interface code is needed to use SQL
queries from a host language for programming non-SQL parts of
applications~\cite{geiger1995inside}

\end{itemize}
Moreover, all these programs are only for computing the transitive closure;
to compute the other three kinds of reachability queries, separate programs
or additional code are needed, but additional code on top of the transitive
closure code will not give the better performance possible for those
queries.

\mypar{Challenges of using rules with everything else for other tasks}
Using rules makes complex reasoning and queries easy, but other language
constructs are needed for other aspects of real-world applications.  How
can one integrate rules with everything else without extra interfaces?
There are several main challenges, for even very basic questions.

\begin{itemize}
\item The most basic question is, how do predicates and logic variables in
  logic languages relate to constructs in commonly-used languages?

  A well-accepted correspondence is: a predicate corresponds exactly to a
  Boolean-valued function, evaluating to true or false on given arguments.
  Indeed, in dominant logic programming languages based on
  Prolog~\cite{Sterling:Shapiro:94}, a predicate defined using rules is
  often used like a function or even procedure (when it uses non-declarative features, such as input and output) defined using expressions
  and statements in commonly-used languages.

  However, a big difference is that, instead of following the control flow from
  function or procedure arguments to the return value or result, techniques
  like unification are used to equate different occurrences of a variable
  in a rule.  Indeed, logic variables in rules are very different from
  variables in commonly-used languages: the former equate or relate
  arguments of predicates, whereas the latter store computed values of
  expressions.

  Thus, the correspondence between predicates and functions, and between
  logic variables and variables in commonly-used languages, is not proper.
  A seamlessly integrated language must establish the proper
  correspondence.

\item Another basic puzzle is even within logic languages themselves.  How
  can a set of rules defined over a particular predicate be used over a
  different predicate?  For example, how can rules defining \co{path} over
  predicate \co{edge} be used over another predicate, say \co{link}?

  Supporting such uses has required higher-order
  predicates or more complex features, e.g.,~\cite{CheKW93}.  They incur
  additional baggage that compromise the ease and clarity of using rules.
  Moreover, there have not been commonly accepted constructs for them.

\item A number of other basic but important language features do not have
  commonly accepted constructs in logic languages, or are not supported at
  all: updates, classes and modules, and concurrency with threads and
  processes~\cite{maier18hist-wbook}.

  Even basic declarative features are often only partially supported in
  logic languages and do not have commonly accepted constructs: 
  set comprehension, aggregation (such as count, sum, or max), and
  general quantification.

  Even the semantics of recursive rules, when simple operations such as
  negation and aggregation are also used,
  has been a matter of significant
  disagreement~\cite{trusz18sem-wbook,gelfond2019vicious,LiuSto20Founded-JLC,LiuSto22RuleAgg-LFCS}.

  A %
  seamlessly integrated language that supports rules must not 
  create complications for using everything else in commonly-used languages.
\end{itemize}

We address these challenges with the Alda language, with an integrated
declarative and operational semantics, allowing complex queries to be
written declaratively, easily and clearly, and be implemented with
generally good performance.

\section{Alda language} 
\label{sec-lang}

We first introduce rules and then describe how our overall language
supports rules with sets and functions as well as imperative updates and
object-oriented programming.  
Figure~\ref{fig-prog} shows an example program in Alda that uses all of rules, sets, functions, updates, and objects.
It will be explained throughout Sections~\ref{sec-rule}--\ref{sec-obj} when used as examples.
Section~\ref{sec-formal} provides an overview
of the formal semantics.  A complete\arxiv{ exposition of the abstract
  syntax and} formal semantics is in
\arxiv{Appendix~\ref{app-formal}}\oops{}.

\begin{figure}[t]
\centering
\begin{oopsenv}
\begin{lstlisting}[numbers=left]
class CoreRBAC:                  # Core RBAC component
  def setup():                   # set up the component
    self.USERS, self.ROLES, self.UR := {},{},{}
                                 # sets of users, roles,
                                 # user-role pairs
  def AddRole(role):             # add role to ROLES
    ROLES.add(role)
  def AssignedUsers(role):       # set of users who have 
    return {u: u in USERS | (u,role) in UR} # the given role
\end{lstlisting}
\end{oopsenv}
\begin{arxivenv}
\begin{lstlisting}[numbers=left]
class CoreRBAC:                  # Core RBAC component
  def setup():                   # set up the component
    self.USERS, self.ROLES, self.UR := {},{},{}
                                 # sets of users, roles, user-role pairs
  def AddRole(role):             # add role to ROLES
    ROLES.add(role)
  def AssignedUsers(role):       # set of users who have the given role 
    return {u: u in USERS | (u,role) in UR}
\end{lstlisting}
\end{arxivenv}
\Vex{-2}
\begin{lstlisting}
  ...
\end{lstlisting}
\Vex{-2}
\begin{oopsenv}
\begin{lstlisting}[numbers=left,firstnumber=8]
class HierRBAC extends CoreRBAC: # Hierarchical RBAC component
  def setup():
    super().setup() # set up sets defined in CoreRBAC
    self.RH := {}   # set of ascendant-descendant role pairs
  def AddInheritance(a,d):       # add (a,d) to RH
    RH.add((a,d))
  rules trans_rs:                # rule set for trans. clo.
    path(x,y) if edge(x,y)
    path(x,y) if edge(x,z), path(z,y)
  def transRH():                 # use infer plus set query
    return infer(path, edge=RH, rules=trans_rs)
           + {(r,r): r in ROLES}
  def AuthorizedUsers(role):     # users who have the role 
                                 # transitively
    return {u: u in USERS, r in ROLES 
             | (u,r) in UR and (r,role) in transRH()}
\end{lstlisting}
\end{oopsenv}
\begin{arxivenv}
\begin{lstlisting}[numbers=left,firstnumber=9]
class HierRBAC extends CoreRBAC: # Hierarchical RBAC extends Core RBAC
  def setup():
    super().setup()              # set up sets defined in CoreRBAC
    self.RH := {}                # set of ascendant-descendant role pairs
  def AddInheritance(a,d):       # add (a,d) to RH
    RH.add((a,d))
  rules trans_rs:                # rule set defining transitive closure
    path(x,y) if edge(x,y)
    path(x,y) if edge(x,z), path(z,y)
  def transRH():                 # transitive RH plus reflexive role pairs
    return infer(path, edge=RH, rules=trans_rs) + {(r,r): r in ROLES}
  def AuthorizedUsers(role):     # users who have the role transitively
    return {u: u in USERS, r in ROLES | (u,r) in UR, (r,role) in transRH()}
\end{lstlisting}
\end{arxivenv}
\Vex{-2}
\begin{lstlisting}
  ...
\end{lstlisting}
\Vex{-2}
\begin{lstlisting}[numbers=left,firstnumber=22]
h = new(HierRBAC)                # create a HierRBAC object
h.AddRole('chair')               # add role 'chair'
\end{lstlisting}
\Vex{-2}
\begin{lstlisting}
...
\end{lstlisting}
\Vex{-2}
\begin{lstlisting}[numbers=left,firstnumber=24]
h.AuthorizedUser('chair')        # query authorized users of role `chair'
\end{lstlisting}
\Vex{-2}
\begin{lstlisting}
...
\end{lstlisting}
\Vex{-2}
\caption{An example program in Alda demonstrating logic rules used with sets, functions, updates, and objects.}
\label{fig-prog}
\end{figure}

\mysubsec{Logic rules}
\label{sec-rule}

We support rule sets of the following form, where
\p{name} is the name of the rule set, \p{declarations} is a set of
predicate declarations, and the body is a set of rules.
\begin{code}
  \RULES \p{name} (\p{declarations}): 
    \p{rule}+
\end{code}
A \defn{rule} is either one of the two equivalent forms below, meaning
that if \p{hypothesis\sb{1}} through \p{hypothesis\sb{h}} all hold, then
\p{conclusion} holds.
\begin{code}
  \p{conclusion} \cIF \p{hypothesis\sb{1}}, \p{hypothesis\sb{2}}, \p{...}, \p{hypothesis\sb{h}}
  \cIF \p{hypothesis\sb{1}}, \p{hypothesis\sb{2}}, \p{...}, \p{hypothesis\sb{h}}: \p{conclusion}
\end{code}
If a conclusion holds without a hypothesis, %
then \co{if} and \co{:} are omitted.

Declarations are about predicates used in the rule set, for advanced uses, and are optional.
For example, they may specify 
argument types of predicates, so rules can be compiled to efficient
standalone imperative
programs~\cite{LiuSto03Rules-PPDP,LiuSto09Rules-TOPLAS} that are expressed
in typed languages~\cite{RotLiu07Retrieval-PEPM}.
We omit the details because they are orthogonal to the focus of the
paper.  In particular, we omit types to avoid unnecessary clutter in code.

We use Datalog rules~\cite{AbiHulVia95,maier18hist-wbook} in examples,
but our method of integrating semantics applies to rules in general. 
Each hypothesis and conclusion in a rule is an \defn{assertion},
of the form\Vex{-.5}
\[\co{\p{p}(\p{arg\sb{1}},\p{...},\p{arg\sb{a}})}\] 
where \p{p} is a \defn{predicate}, and each \p{arg_k} is a variable 
or a constant.
We use numbers and quoted strings to represent constants, and the rest are
variables.
As is standard for safe rules, all variables in the conclusion must be in a hypothesis.
If a conclusion holds without a hypothesis, %
then each argument in the conclusion must be a constant, in which case the
conclusion is called a \defn{fact}.
Note that a predicate is also called a \defn{relation}, relating the
arguments of the predicate.

\begin{example}
  For computing the transitive closure of a graph in the running example,
  the rule set, named \co{trans\_rs}, in Figure~\ref{fig-prog} (lines 15-17) can be written.
  The rules are the same as in dominant logic languages except for
  the use of lower-case variable names, the change of \co{:-} to \co{if}, and
  the omission of dot at the end of each rule.
\end{example}

\mypar{Terminology}
In a rule set, predicates not in any conclusion are called \defn{base
  predicates} of the rule set, and the other predicates are called
\defn{derived predicates} of the rule set.

We say that a predicate \p{p} \defn{depends on} a predicate \p{q} if \p{p}
is in the conclusion of a rule whose hypotheses contain \p{q} or contain a
predicate that depends on \p{q} recursively.

We say that a derived predicate \p{p} in a rule set \p{rs} \defn{fully
  depends on} a set \p{s} of base predicates in \p{rs} if \p{p} does not
depend on other base predicates in \p{rs}.

\begin{example}
  In rule set \co{trans\_rs}, \co{edge} is a base predicate, and
  \co{path} is a derived predicate.  \co{path} depends on \co{edge} and
  itself.  \co{path} fully depends on \co{edge}.
\end{example}

\mysubsec{Integrating rules with sets, functions, updates, and objects}
\label{sec-integrate}

Our overall %
language supports all of rule sets and the following language constructs as
built-ins; all of them can appear in any scope---global, class, and local.
\begin{itemize}

\item Sets and set expressions (comprehension, aggregation, quantification,
  and high-level operations such as union) to make non-recursive queries
  over sets easy to express.

\item Function and procedure definitions with optional keyword arguments,
  and function and procedure calls.

\item Imperative updates by assignments and membership\oops{} 
  changes, to sets and
  other data, in sequencing, branching, and looping statements.

\item Class definitions containing object field and method (function and procedure) definitions, object creations, and inheritance.

\end{itemize}
A name holding any value is \defn{global} if it is introduced (declared or
defined) at the global scope; is an \defn{object field} if it is introduced
for that object;
or is \defn{local} to the function, method, or rule set that contains it
otherwise.
The value that a name is holding is available after the name is defined:
globally for a global name, on the object for an object field, and in the
enclosing function, method, or rule set for a local name.
\notes{}

\begin{example}
  Rule set \co{trans\_rs} in Figure~\ref{fig-prog} (defined on line 15 and queried on line 19) is used together with sets (defined on lines 3 and 12), set expressions (on lines 8, 19, and 21), functions (defined on lines 7, 18, and 20), procedures (defined on lines 2, 5, 10, and 13), updates (on lines 3, 6, 12, 14), classes (defined on lines 1 and 9, with inheritance), and objects (created on line 22).
  No extra code is needed to convert \co{edge} and \co{path}, declare logic variables, and so on.
\end{example}

The key ideas of our seamless integration of rules with sets, functions,
updates, and objects are:
(1) a predicate is a set-valued variable that holds a set of tuples for
which the predicate is true,
(2) queries using rules are calls to an inference function that computes
desired sets using given sets,
(3) values of predicates can be updated either directly as for other
variables or by the inference function, and
(4) predicates and rule sets can be object attributes %
as well as global and local names, just as sets and functions can.

\mypar{Integrated semantics, ensuring declarative semantics of rules}
In our overall language, the meaning of a rule set \p{rs} is completely
declarative, exactly following the standard least fixed-point semantics of
rules~\cite{fitting2002fixpoint,LiuSto09Rules-TOPLAS}:
\begin{itemize}
\item[] Given values of any set \p{s} of base predicates in \p{rs}, the
  meaning of \p{rs} is, for all derived predicates in \p{rs} that fully
  depend on \p{s}, the least set of values that can be inferred, directly
  or indirectly, by using the given values and the rules in \p{rs};

  for any derived predicate in \p{rs} that does not fully depend on
  \p{s},
  i.e., depends on any base predicate whose values are not given, its value
  is \defn{undefined}.
\end{itemize}
The operational semantics for the rest of the language ensures this
declarative semantics of rules.  

The precise constructs for using rules with sets, functions, updates, and
objects are described in Sections~\ref{sec-pred}--\ref{sec-obj}.

\mysubsec{Predicates as set-valued variables}
\label{sec-pred}
For rules to be easily used
with everything else, our most basic principle in 
designing the language
is to treat a predicate as a set-valued variable that holds the set of tuples
that are true for the predicate, that is:
\begin{quote} %
  For any predicate \p{p} over values \co{\p{x\sb{1}},\p{...},\p{x\sb{a}}},
  assertion \co{\p{p}(\p{x\sb{1}},\p{...},\p{x\sb{a}})} is true---i.e.,
  \co{\p{p}(\p{x\sb{1}},\p{...},\p{x\sb{a}})} is a fact---if and only if
  tuple \co{(\p{x\sb{1}},\p{...},\p{x\sb{a}})} is in set \co{\p{p}}.
  Formally,\Vex{-1}
\[
\co{\p{p}(\p{x\sb{1}},\p{...},\p{x\sb{a}})} 
~\Longleftrightarrow~ %
\co{(\p{x\sb{1}},\p{...},\p{x\sb{a}}) \IN \p{p}}
\]
\end{quote}
This means that, as variables,
predicates in a rule set can be introduced in any scope---as 
global variables, 
object fields, %
or variables local to the rule set---and 
they can be written into and read from without needing any extra
interface.

\begin{example}
  In rule set \co{trans\_rs} in Figure~\ref{fig-prog}, predicate \co{edge}
  is exactly a variable holding a set of pairs, such that
  \co{edge(\m{x},\m{y})} is true iff \co{(\m{x},\m{y})} is in \co{edge}, and \co{edge} is local to \co{trans\_rs}
  In general, \co{edge} can be a global variable,
  an object field, or a local variable of \co{trans\_rs}.
  Similarly for predicate \co{path}.
\end{example}

Writing to predicates is discussed later under updates to predicates, but
reading and using values of predicates can simply use all operations on
sets.
We use set expressions including the following:
\Vex{1}\\
{\oops{}
\begin{tabular}{@{\Hex{2}}l@{\Hex{1}\arxiv{\Hex{10}}}l}
    \co{\p{exp} \IN \p{sexp}}
    & \Hex{-0}membership\\
    \co{\p{exp} \NOT \IN \p{sexp}}
    & \Hex{-0}negated membership\\
    \co{\p{sexp\sb{1}} + \p{sexp\sb{2}}}
    & \Hex{-0}union\\
    \co{\{\p{exp}:\,\p{v\sb{1}} \IN \p{sexp\sb{1}},\p{...},\p{v\sb{k}} \IN \p{sexp\sb{k}}\,|\,\p{bexp}\}}
    & comprehension\\
\arxiv{
    \co{\p{agg} \p{sexp}}, \Hex{1} where \co{\p{agg}} is \co{count}, \co{max}, \co{min}, \co{sum}
    & aggregation \\
}
    \co{\SOME~\p{v\sb{1}} \IN \p{sexp\sb{1}},\p{...},\p{v\sb{k}} \IN \p{sexp\sb{k}}\,|\,\p{bexp}}
    & existential quantification
\end{tabular}
}\Vex{1}\\
A comprehension returns the set of values of 
\p{exp} for all combinations of values of variables that satisfy
all membership clauses \co{\p{v_i} \IN \p{sexp_i}} and condition 
\p{bexp}.
\arxiv{An aggregation returns the count, max, etc. of the set value of \p{sexp}.}
An existential quantification returns true iff for some %
combination of values of variables that satisfies all \co{\p{v_i} \IN \p{sexp}} clauses, condition \p{bexp} holds. When an existential quantification returns true, variables \p{v_1},...,\p{v_k} are bound to a witness.
Note that these set queries, as in~\cite{Liu+17DistPL-TOPLAS}, are more powerful than those in Python.

\begin{example}
  For computing the transitive closure \co{T} of a set \co{E} of edges, the
  following \co{\WHILE} loop with quantification can be used (we will see
  that we use objects and updates as in Python except the syntax \co{:=}
  for assignment in this paper):
\begin{lstlisting}
  T := E.copy()
  while some (x,z) in T, (z,y) in E | (x,y) not in T:
    T.add((x,y))
\end{lstlisting}\oops{}\arxiv{
This is simpler than the Python \co{\WHILE} loop in Section~\ref{sec-prob}:
it finds a witness pair \co{(x,y)} directly using \co{some}, instead of 
constructing workset \co{W} and then using \co{pop} to get a pair.}
\end{example}

In the comprehension and aggregation forms, each \p{v_i} can also be a
tuple pattern that elements of the set value of \p{sexp_i} must
match~\cite{Liu+17DistPL-TOPLAS}.
A \defn{tuple pattern} is %
a tuple in which each component is a non-variable expression,
a variable possibly prefixed with \co{=}, a wildcard \co{\_}, or
recursively a tuple pattern.
For a value to match a tuple pattern, 
it must have the corresponding tuple structure, with corresponding
components equal the values of non-variable expressions and variables
prefixed with \co{=}, and with corresponding components assigned to
variables not prefixed with \co{=}; corresponding components for wildcard
are ignored.

\begin{example}
  To return the set of second component of pairs in \co{path} whose first
  component equals the value of variable \co{x}\arxiv{, and where that second
  component is also the first component of pairs in \co{edge} whose second
  component is 1}, one may use a set comprehension with tuple patterns:
\begin{oopsenv}
\begin{lstlisting}
  {y: (=x,y) in path}
\end{lstlisting}
\end{oopsenv}
\begin{arxivenv}
\begin{lstlisting}
  {y: (=x,y) in path, (y,1) in edge}
\end{lstlisting}\Vex{-3}
\end{arxivenv}
\end{example}

Now that predicates in rules correspond to set-valued variables, instead of
functions or procedures, we can further see that\arxiv{ logic variables,
  i.e.,} variables in rules, are like\arxiv{ pattern variables, i.e.,} 
variables not prefixed with \co{=} in patterns.
Logic rules do not have variables that store values, i.e., variables
prefixed with \co{=} in patterns.

\mysubsec{Queries as calls to an inference function} %
\label{sec-infer}
For inference and queries using rules, calls to a built-in inference
function \co{\INFER}, of the following form, are used, with
\p{query\sb{k}}'s and \co{\p{p\sb{k}}=\p{sexp\sb{k}}}'s being optional:\m{\!}
\begin{code}
  \INFER(\p{query\sb{1}},\,\p{...},\,\p{query\sb{j}}, \p{p\sb{1}}=\p{sexp\sb{1}},\,\p{...},\,\p{p\sb{i}}=\p{sexp\sb{i}}, \RULES=\p{rs})
\end{code}
\co{\p{rs}} is the name of a rule set.  Each \co{\p{sexp\sb{k}}} is a
set-valued expression.  Each \co{\p{p\sb{k}}} is a base predicate of \p{rs}
and is local to \p{rs}.
Each \co{\p{query\sb{k}}} is of the form
\co{\p{p}(\p{arg_1},\p{...},\p{arg_a})},
where \co{\p{p}} is a derived predicate of \p{rs}, and each %
argument \co{\p{arg_k}} is a constant, 
a variable possibly prefixed with
\co{=}, or wildcard \co{\_}.
A variable prefixed with \co{=} indicates a bound variable whose value will
be used as a constant when evaluating the query.
\arxiv{So arguments of queries are patterns too.}
If all \co{\p{arg_k}}'s are \co{\_}, the abbreviated form \co{\p{p}} 
can be used.

Function \co{\INFER} can be called implicitly by the language implementation
or explicitly by the user.
It is called automatically as needed and can be called explicitly when desired.

\begin{example} 
  For inference using rule set \co{trans\_rs} in Figure~\ref{fig-prog}, where
  \co{edge} and \co{path} are local variables, \co{\INFER} can be called in
  many ways, including: %
\begin{lstlisting}
  infer(path, edge=RH, rules=trans_rs)
  infer(path(_,_), edge=RH, rules=trans_rs)
  infer(path(1,_), path(_,=R), edge=RH, rules=trans_rs)
\end{lstlisting} %
The first is as in Figure~\ref{fig-prog} (line 19).
The first two calls are equivalent: \co{path} and \co{path(\_,\_)} both
query the set of pairs of vertices having a path from the first vertex to
the second vertex, 
following edges given by the value of variable \co{RH}.
In the third call, \co{path(1,\_)} queries the set of vertices having a
path from vertex 1, and \co{path(\_,=R)} queries the set of vertices having
a path to the vertex that is the value of variable \co{R}.

If \co{edge} or \co{path} is a global variable or an object field, 
one may call \co{\INFER} on \co{trans\_rs} without 
assigning to \co{edge} or querying %
\co{path}, respectively.
\end{example}

The operational semantics of a call to \co{\INFER}
is exactly like other function calls, except for the special forms of
arguments and return values, and of course the inference function performed
inside:
\begin{enumerate}

\item[1)] For each value \p{k} from 1 to \p{i}, assign the set value of
  expression \p{sexp_k} to predicate \p{p_k} that is a base predicate of
  rule set \p{rs}.

\item[2)] Perform inference using the rules in \p{rs} and the given values
  of base predicates of \p{rs} following the declarative semantics,
  including assigning to derived predicates that are not local.

\item[3)] For each value \p{k} from 1 to \p{j},
return the result of query \p{query_k} as the \p{k}th component of the
return value.
The result of a query with \m{l} distinct variables not prefixed with \co{=} is a set of tuples of
\m{l} components, one for each of the distinct variables in their order of
first occurrence in the query.
\end{enumerate}
Note that
when there are no \co{\p{p\sb{k}}=\p{sexp\sb{k}}}'s, only defined values of
base predicates that are not local to \p{rs} are used; and
when there are no \p{query\sb{k}}'s, only values of derived predicates that
are not local to \p{rs} may be inferred and no value is returned.

\arxiv{
\arxiv{Section~\ref{sec-rbac}}\oops{} on
benchmarks using Role-Based Access Control (RBAC) discusses different ways
of using rules and different
calls to \co{\INFER}: implicit vs.\ explicit, in an enclosing expression
vs.\ by itself, passing in all base predicates vs.\ only some, etc. %
}

\mysubsec{Updates to predicates}
\label{sec-upd}
Values of base predicates can be updated directly as for other set-valued
variables, and values of derived predicates are updated by the inference
function.

Base predicates of a rule set \p{rs} that are local to \p{rs} are assigned
values at calls to \co{\INFER} on \p{rs}, as described earlier.  Base
predicates that are not local can be updated by using assignment statements
or set membership update operations.  
We use
\begin{code}
  \p{lexp} := \p{exp}
\end{code} 
for assignments, where \co{\p{lexp}} can
also be a nested tuple of variables, and each variable
is assigned the corresponding component of the value of \co{\p{exp}}. 

Derived predicates of a rule set \p{rs} can be updated only by calls to the
inference function on \p{rs}.  These updates must ensure the declarative
semantics of \p{rs}:
\begin{itemize}

\item[] Whenever a base predicate of \p{rs} is updated in the program, the
  values of the derived predicates in \p{rs} are maintained
  according to the declarative semantics of \p{rs} by
  calling \co{\INFER} on \p{rs}.

  Updates to derived predicates of \p{rs} outside \p{rs} are not allowed,
  and any violation will be detected and reported at compile time if
  possible and at runtime otherwise.

\end{itemize}
Simply put, updates to base predicates trigger updates to derived
predicates, and other updates to derived predicates are not allowed.
This ensures the invariants that the derived predicates hold the values
defined by the rule set based on values of the base
predicates, as required by the declarative semantics.
Note that this is the most straightforward semantics, but the
implementation can avoid many inefficiencies with optimizations\arxiv{, as described in
\arxiv{Section~\ref{sec-optimize}}\oops{}}.

\begin{example}
  Consider rule set \co{trans\_rs} in Figure~\ref{fig-prog}. 
  If \co{edge} is not local, one may assign a set of pairs to \co{edge}:
\begin{lstlisting}
  edge := {(1,8),(2,9),(1,2)}
\end{lstlisting}

If \co{edge} is local, the example calls to \co{\INFER} in
Section~\ref{sec-infer} assign the value of \co{RH} to \co{edge}.

If \co{path} is not local, then a call \co{\INFER(edge=RH, \RULES=trans\_rs)}
updates \co{path}, contrasting the first two calls to \co{\INFER} in the
previous example that return the value of \co{path}.

If \co{path} is local, the return value of \co{\INFER} can be assigned to
variables.  For example, for the third example call to \co{\INFER} in
Section~\ref{sec-infer}, this can be
\begin{oopsenv}
\begin{lstlisting}
  from1,toR := infer(path(1,_), path(_,=R), edge=RH, 
                     rules=trans_rs)
\end{lstlisting}
\end{oopsenv}
\begin{arxivenv}
\begin{lstlisting}
  from1,toR := infer(path(1,_), path(_,=R), edge=RH, rules=trans_rs)
\end{lstlisting}
\end{arxivenv}

If both \co{edge} and \co{path} are not local, then whenever \co{edge} is
updated,
an implicit call \co{\INFER(\RULES=trans\_rs)} is executed automatically to
update the value of \co{path}.
\end{example}

\mysubsec{\oops{}Using predicates and rules with objects and classes} 
\label{sec-obj}
Predicates and rule sets can be object fields as well as global and local
names, just as sets and functions can, as discussed in
Section~\ref{sec-integrate}.  This allows predicates and rule sets to be
used seamlessly with objects
in object-oriented programming. %

For other constructs than those described above, we use those in high-level
object-oriented languages.  We mostly use Python syntax (looping,
branching, indentation for scoping, `\co{:}' for elaboration, `\co{\#}' for
comments, etc.) for succinctness, but with a few conventions from Java
(keyword \co{new} for object creation, 
keyword \co{extends} for subclassing,
and omission of \co{self}, the equivalent of \co{this} in Java, 
when there is no ambiguity)
for ease of reading.

\begin{example}
We use %
Role-Based Access Control (RBAC) to show the need of using rules with all
of sets, functions, updates, and objects and classes.

RBAC is a security policy framework for controlling user access to
resources based on roles 
and is widely used in large organizations.  The ANSI standard for
RBAC~\cite{ansi04role} was approved in 2004 after several 
rounds of public
review~\cite{Sandhu+00,Jaeger:Tidswell:00,ferraiolo01proposed}, building on
much research during the preceding decade and earlier.
High-level executable specifications were developed for the entire RBAC
standard~\cite{LiuSto07RBAC-ONR}, where all queries are declarative except
for computing the transitive role-hierarchy relation in Hierarchical RBAC,
which extends Core RBAC.

Core RBAC defines functionalities relating users, roles, permissions, and
sessions. 
It includes the sets and update and query functions
in class \co{CoreRBAC} in Figure~\ref{fig-prog}, as in~\cite{LiuSto07RBAC-ONR}\footnote{Only a few selected sets and
  functions are included, and with small changes to names and syntax.}.

Hierarchical RBAC adds support for a role hierarchy, \co{RH}, and update
and query functions extended for \co{RH}.  It includes the update and query
functions in class \co{HierRBAC} in Figure~\ref{fig-prog}, as in~\cite{LiuSto07RBAC-ONR}\m{^1}, except that function \co{transRH()} in~\cite{LiuSto07RBAC-ONR} computes the
transitive closure of \co{RH} plus reflexive role pairs for all roles in
\co{ROLES} by using a complex and inefficient \co{\WHILE} loop
similar to that 
in Section~\ref{sec-prob} plus a union with the set of reflexive role pairs
\co{\{(r,r):~r in ROLES\}}, whereas function \co{transRH()} in Figure~\ref{fig-prog} simply calls \co{\INFER} and unions the result with
reflexive role pairs.

Note though, in the RBAC standard, a relation \co{transRH} is used in place
of \co{transRH()}, intending to maintain the transitive role hierarchy
incrementally while \co{RH} and \co{ROLES} change.  It is believed that this is done
for efficiency, because the result of \co{transRH()} is used continually,
while \co{RH} and \co{ROLES} change infrequently.  However, the maintenance
was done inappropriately~\cite{LiuSto07RBAC-ONR,Li+07critique} and
warranted the use of \co{transRH()} to ensure correctness before
efficiency.

Overall, the RBAC specification relies extensively on all of updates, sets,
functions, and objects and classes with inheritance, besides rules:
(1) updates for setting up and updating the state of the RBAC system,
(2) sets and set expressions for holding the system state and expressing
set queries exactly as specified in the RBAC standard,
(3) methods and functions for defining and invoking update and query
operations\arxiv{, including
function \co{transRH()}}, and
(4) objects and classes for capturing different components---\co{CoreRBAC}, \co{HierRBAC}, constraint RBAC, their further refinement, extensions, and
combinations, totaling 9 components, corresponding to 9 classes, including
5 subclasses of \co{HierRBAC}~\cite{ansi04role,LiuSto07RBAC-ONR}.
\end{example}

\subsection{Formal semantics}
\label{sec-formal}

Formal semantics of logic rules has been studied
extensively, including the
standard least fixed-point semantics for Datalog and more~\cite{fitting2002fixpoint,LiuSto20Founded-JLC}. 
A formal operational semantics for DistAlgo, a powerful language with all
of sets, functions, updates, and objects, including even distributed
processes as objects, but without rules, has also been given
recently~\cite{Liu+17DistPL-TOPLAS}.

Building on these prior semantics,
we developed a formal semantics for a core language for Alda that preserves the semantics from all above and seamlessly connects declarative rule semantics and imperative update semantics.  We removed the constructs specific to distributed processes and added the constructs described in this paper.  The removed DistAlgo constructs can easily be restored to obtain a semantics for the full language; we removed them simply to avoid repeating them.

\oops{The Supplementary Material}\arxiv{Appendix~\ref{app-formal}}
contains details of the abstract syntax and semantics for our core language.  This section presents a brief high-level overview of the semantics. 
  


The operational semantics is a reduction semantics with evaluation contexts
\cite{wright94syntactic,serbanuta07rewriting}.  It culminates in the
definition of a transition relation between states.  A state has the form
$\langle s,ht,h,{\it stk}\rangle$.  $s$ is the statement to be
executed. $ht$ is the heap type map: $ht(a)$ is the type of the object on
the heap at address $a$.  $h$ is the heap; it maps addresses to objects.
${\it stk}$ is a special call stack used to track the rule sets whose
results should be automatically maintained.  It is initialized with an
entry containing rule sets defined in global scope.  When a method is
called on an instance at address $a$ of a user-defined class $c$, the call
stack is extended by pushing an entry containing the rule sets defined in
$c$, instantiated by replacing {\tt self} with $a$; the entry is popped
when the method returns.

The transition relation for statements has the form $\sigma \rightarrow \sigma'$ where $\sigma$ and $\sigma'$ are states.  It is implicitly parameterized by the program.  The transition rules for assignment statements and calls to set membership update operations (e.g., \co{add}), user-defined methods, and \co{\INFER} 
are extended to maintain the results of all rule sets on the call stack.  We outline two of these transition rules here as representative examples.

The transition rule for calling a method on an instance at address $a$ of a user-defined class replaces the method call with a copy of the method body instantiated by substituting argument values for parameters, 
pushes onto the call stack an entry containing instantiated rule sets as
described above, and updates the heap type map and heap to
capture the results of automatic maintenance using %
all rule sets on the call stack.  Automatic maintenance performs inference for each of those rule sets using values of their base predicates in the current state $\sigma$, and then updates the values of all their derived predicates in the next state $\sigma'$, like a single parallel-assignment statement.  Use of parallel assignment is significant, because a derived predicate of one rule set can be a base predicate of another rule set.  Values of predicates are instances of the built-in \co{set} class and are stored on the heap, just like other objects.

The transition rule for an explicit call to \co{\INFER} on a rule set
instantiates that rule set using the given values for the rule set's
parameters, updates the heap type map and heap to capture the results of
evaluating the instantiated rule set and returning the query results, and then updates the heap type map and heap to capture
the results of automatic maintenance using all rule sets on the call
stack, as described above.


\Vex{-.5}
\section{Compilation and optimization}
\label{sec-impl}

The operational semantics to ensure the declarative semantics of rules is
conceptually simple, but the implementation required can vary widely,
depending on the kinds of updates
in the programs.  It can be extremely expensive or complex when there are
arbitrary updates in an object-oriented language that allows aliasing of
object references.

We first present how to compile all possible updates to predicates,
starting with the checks and actions needed for correctly handling updates
for a single rule set with implicit and explicit calls to \co{\INFER}.
We then describe how to implement the inference in \co{\INFER}.
Additionally\oops{}, we systematize powerful optimizations that can be exploited in the overall compilation framework.

\Vex{-.5}
\mysubsec{Compiling %
updates to predicates} %

The implementation required by the operational semantics of a rule set
\p{rs} falls into
three cases, based on the nature of updates to base predicates of \p{rs}
outside \p{rs}.  Note that inside \p{rs} there are no updates to base predicates of
\p{rs}, by definition of base predicate.
\begin{enumerate}
\item[1)] {\bf Non-local updates with aliasing.} For updates to non-local
  variables of \p{rs} 
  in the presence of variable aliasing (i.e., two different variables
  referring to the same value or object), each update needs to check
  whether the variable updated may alias a base predicate of \p{rs} and, if
  so, an implicit call to \co{\INFER} on \p{rs} needs to be made.

  An update outside \p{rs} to a non-local variable that is a derived
  predicate of \p{rs} needs to be detected and reported, conservatively at
  compile-time if possible, and at runtime otherwise.
  Note that this also solves the most nasty problem of possible
  aliasing between a derived predicate and a base predicate, because
  updating a base predicate of \p{rs} outside \p{rs} would also be updating
  a derived predicate of \p{rs} outside \p{rs}, which would be detected and
  reported.

\item[2)] {\bf Non-local updates without aliasing.} For updates to non-local
  variables of \p{rs} when there is no variable aliasing, an implicit call
  to \co{\INFER} on \p{rs} needs to be made only after every update to a
  base predicate of \p{rs}.

  Without aliasing, statements outside \p{rs} that update derived
  predicates of \p{rs} can be identified and reported as errors at compile
  time.

\item[3)] {\bf Local updates by explicit calls.} Local variables of \p{rs}
  can be assigned values only at explicit calls to \co{\INFER} on \p{rs}.
  Such a call passes in values of local variables that are base predicates
  of \p{rs} before doing the inference.
  Values of local variables that are derived predicates of \p{rs} can only
  be used in constructing answers to the queries in the call, and the
  answers are returned from the call.

  There are no updates outside \p{rs} to local variables that are 
  derived predicates of \p{rs}, %
  by definition of local variables.

\end{enumerate}
To satisfy these requirements, the overall method for compiling an
update to a variable \co{v} outside rule sets %
is:
\begin{itemize}

\item In the presence of aliasing: insert code that does the following
  after the update:
  if \co{v} refers to a derived predicate of any rule set, report a runtime
  error and exit; otherwise for each rule set \p{rs}, if \co{v} refers to
  a base predicate of \p{rs}, call \co{\INFER} on \p{rs} with no arguments
  for base predicates and no queries.

\item In the absence of aliasing: report a compile-time error if \co{v} is
  a derived predicate of any rule set; otherwise, for each rule set \p{rs}
  that contains \co{v} as a base predicate, insert code, after the update,
  that calls \co{\INFER} on \p{rs} with no arguments for base predicates and no
  queries.

\end{itemize}
Compiling an explicit call to \co{\INFER} on a rule set
directly follows the operational semantics of \co{\INFER}.

In effect, function \co{\INFER} %
is called to implement a wide range of control: from inferring everything
possible using all rule sets and values of all base predicates at every
update in the most extensive case, to answering specific queries using specific rules and specific
sets of values of specific base predicates at explicit calls.

Obviously, use of updates %
in different contexts has significant impact on program efficiency.  It is
particularly worth noting that the very existence of aliasing, intended to
provide efficiency at the cost of tedious and error-prone programming,
incurs the most performance penalty.

\Vex{-.5}
\mysubsec{Implementing inference and queries} %

Any existing method can be used to implement the functionality inside
\co{\INFER}.  The inference and queries for a rule set can use either
bottom-up or top-down
evaluation~\cite{KifLiu18wbook,TekLiu10RuleQuery-PPDP,TekLiu11RuleQueryBeat-SIGMOD},
so long as they use the rule set and values of the base
predicates according to the declarative semantics of rules

The inference and queries can be either performed by using a general logic
rule engine, e.g., XSB~\cite{swift2012xsb,xsb21}, 
or compiled to specialized standalone executable code as in,
e.g.,~\cite{LiuSto09Rules-TOPLAS,RotLiu07Retrieval-PEPM}, that is then
executed.
\arxiv{Our implementation uses the former approach,
  by indeed using the well-known XSB system, 
  as described in Section~\ref{sec-expe}, because it allows easier
  extensions to support more kinds of rules and optimizations that are
  already supported in XSB.}

\notes{} %

\arxiv{\oops{\section}\arxiv{\mysubsec}{Powerful optimizations}
\label{sec-optimize}
Efficient inference and queries using rules is well known to be challenging
in general, and especially so if it is done repeatedly to ensure the
declarative semantics of rules under updates to predicates.  Addressing the
challenges has produced an extensive literature in several main areas in
computer science---database, logic programming, automated reasoning, and
artificial intelligence in general---and is not the topic of this paper.

Here, we describe how well-known analyses and optimizations can be used
together to improve the implementation of the overall language as well as
the rule language, giving a systemic perspective of all main optimizations
for efficient implementations.  There are two main areas of optimizations.

The first area is for inference under updates to the predicates used.
There are three main kinds of optimizations in this area: (1) reducing
inference triggered by updates, (2) performing inference lazily only when
the results are demanded, and (3) doing inference incrementally when
updates must be handled to give results:

\begin{description}
\item[Reducing update checks and inference.]
In the presence of aliasing, it can be extremely inefficient to check, for
all rule sets after every update, that the update is not to a derived
predicate of the rule set and whether a call to \co{\INFER} on the rule set
is needed, not knowing statically whether the update affects a base
predicate of the rule set.
Alias analysis, e.g.,~\cite{Goy05,Gor+10Alias-DLS}, can help reduce such
checks by statically determining updates to variables that possibly alias a
predicate of a rule set.

\item[Demand-driven inference.] 
Calling \co{\INFER} after every update to a base predicate can be inefficient and wasteful,
because updates can occur frequently while the maintained derived
predicates are rarely used.
To avoid this inefficiency, \co{\INFER} can be called on demand just before
a derived predicate is used, e.g.,~\cite{FonUll76,RotLiu08OSQ-GPCE,Liu+16IncOQ-PPDP},
instead of immediately after updates to base predicates.

\item[Incremental inference.] 
More fundamentally, even when derived predicates are frequently used,
\co{\INFER} can easily be called repeatedly on slightly changed or even
unchanged base predicates, in which case computing the results from scratch is extremely wasteful.
Incremental computation can drastically reduce this inefficiency by
maintaining the values of derived predicates
incrementally, e.g.,~\cite{GupMum99maint,SahaRam03}.

\end{description}

The second area is for efficient implementation of  rules by themselves,
without considering updates to the predicates used.  There are two main
groups of optimizations.

\begin{description}
\item[Internal demand-driven and incremental inference.] 
\oops{~~~~\linebreak}
Even in a single call to \co{\INFER}, significant
optimizations are needed. 

In top-down evaluation (which is already driven by the given query as
demand), subqueries can be evaluated repeatedly, so
tabling~\cite{TamSat86,CheWar96} (a special kind of incremental computation
by memoization) is critical for avoiding not only repeated evaluation of
queries but also non-termination when there is recursion.

In bottom-up evaluation (which is already incremental from the ground up),
demand
transformation~\cite{TekLiu10RuleQuery-PPDP,TekLiu11RuleQueryBeat-SIGMOD},
which improves over magic sets~\cite{Ban+86} exponentially, can transform
rules to help avoid computations not needed to answer the given query.

\item[Ordering and indexing for inference.]
Other factors can also drastically affect the performance of logic queries
in a single call to \co{\INFER}~\cite{maier18hist-wbook,liu18LPappl-wbook}.  

Most prominently, in dominant logic rule engines like XSB, changing the
order of joining hypotheses in a rule can impact performance dramatically,
e.g., for the transitive closure example, reversing the two hypotheses in
the recursive rule can cause a linear factor performance difference.
Reordering and indexing~\cite{LiuSto09Rules-TOPLAS,Liu+16IncOQ-PPDP}
are needed to avoid such severe slowdowns.
\end{description}

}

\Vex{-.5}
\section{Programming and performance benchmarks}
\label{sec-bench}

We have used Alda for a %
variety of well-known problems where rules can
be used for both ease of programming and performance of execution.
We describe a set of benchmarks for programming and performance evaluation.
One can see that even for problems that were previously focused on for
using rules, it becomes much easier to program using an integrated language
like Alda.

We developed three sets of benchmarks, from
OpenRuleBench~\cite{Lia+09open}, RBAC~\cite{ansi04role,LiuSto07RBAC-ONR},
and program analysis.
OpenRuleBench benchmarks show the wide range of application problems
previously developed using different kinds of rule systems.
RBAC benchmarks show the use of rules in an application that requires all
of sets, functions, updates, and objects and classes, and show different
ways of using rules.
Program analysis benchmarks demonstrate seamlessly integrated use of rules
with also aggregate queries and recursive functions; we also contrast with
using aggregate queries in rule languages, which are not used in any
benchmark in OpenRuleBench.\oops{}

\mysubsec{OpenRuleBench---a wide variety of rule-based applications}

OpenRuleBench~\cite{Lia+09open} contains a wide variety of database,
knowledge base, and semantic web application problems, written using rules
in 11 well-known rule systems from 5 different categories, as well as large
data sets and a large number of test scripts for running and measuring the
performance.  Among 14 benchmarks described in~\cite{Lia+09open}, we
consider all except for\oops{} one that tests interfaces of
rule systems with databases (which is a non-issue for Alda because it extends
Python which has standard and widely-used database interfaces).

Table~\ref{tab-rulebench} summarizes the benchmarks.  We compare with the
benchmark programs in XSB, for three reasons: (1) XSB has been the most
advanced rule system supporting well-founded semantics for non-stratified
negation and tabling techniques for efficient query evaluation, and has
been actively developed for over three decades, to this day;
(2) among all systems reported in~\cite{Lia+09open}, XSB
was one of the fastest, if not the fastest, and the most consistent across
all benchmarks; and (3) among all measurements reported, only XSB,
OntoBroker, and DLV could run all benchmarks, but OntoBroker went bankrupt,
and measurements for DLV were almost all slower, often by orders of
magnitude.

We easily translated all 13 benchmarks into Alda, automatically for all
except for three cases where the original rules used features beyond
Datalog, which became two cases after we added support for negation.  In
all cases, it was straightforward to express the desired functionality in
Alda, producing a program that is very similar or even simpler.
Additionally, the code for reading data, running tests, timing, and writing
results is drastically simpler in Alda as it extends Python.  These special
cases and additional findings are described below.

\begin{table}[t]
  \arxiv{\small}\oops{}
  \centering
\begin{tabular}{@{}l@{}|@{~}p{\arxiv{60}\oops{}ex}@{~}|@{~}r@{~}|@{~}r@{}}
  Name    & Description & Prog size & Data size \\\hline\hline
  Join1   & non-recursive tree of binary joins as inference rules
                        &  225 & * \\\hline
  Join2   & join from IRIS system producing large intermediate result
                        &   41 & * \\\hline
  JoinDup & join of separate results of five copies of Join1
                        &  163 & * \\\hline
  LUBM    & university database adapted from LUBM benchmark
                        &  377 & * \\\hline
  Mondial & geographical database derived from CIA Factbook
                        &   36 & 59,733\\\hline
  DBLP    & well-known bibliography database on the Web
                        &   20 & 2,437,867\\\hline\hline
  TC      & classical transitive closure of a binary relation
                        &   75 & * \\\hline
  SG      & well-known same-generation siblings problem
                        &   90 & * \\\hline
  WordNet & natural language processing queries based on WordNet
                        &  298 & 465,703\\\hline
  Wine    & well-known OWL wine ontology as rules
                        & 1103 & 654 \\\hline\hline
  ModSG   & modified SG to exclude ancestor-descendant relationships
                        &   38 & * \\\hline\arxiv{
  Win     & well-known win-not-win game with non-stratified negation
                        &   24 & * \\\hline
  MagicSet & non-stratified rules from magic-set transformation
                        &   34 & * \\\hline}%
  \hline
\end{tabular}
\caption{Benchmarks for problems in OpenRuleBench~\cite{Lia+09open}.
  The three groups (of 6, 4, \oops{}\arxiv{3}) in order are called 
  large join tests,
  Datalog recursion, and default negation, respectively.
  Prog size is the XSB program size 
  in lines of code without comments and empty lines.
  Data size is the input data size in number of facts; 
  * means that scripts are used to generate input data of desired sizes.\oops{}}
\label{tab-rulebench}
\end{table}

\mypar{Result set}
In most logic languages, including Prolog and many variants, a query
returns only the first result that matches the query.  To return the set of
all results, some well-known tricks are used.  The LUBM benchmark
includes the following extra rules to return all answers of \co{query9\_1}:
\begin{code}
  query9 :- query9_1(X,Y,Z), fail.
  query9 :- writeln('========query9.======').
\end{code}
The first rule first queries \co{query9\_1} to find an answer (a triple of
values for \co{X},\co{Y},\co{Z}) and then uses \co{fail} to trick the
inference into thinking that it failed to find an answer and so continuing
to search for an answer; and it does this repeatedly, until \co{query9\_1}
does fail to find an answer after exhausting all answers.  The second rule
is necessary, even if with an empty right side, to trick the inference into
thinking that it succeeded, because the first rule always ends in failing;
this is so that the execution can continue to do the remaining work instead
of stopping from failing.

\arxiv{In fact, this trick is used for all benchmarks,
  but other uses are buried inside the code for running, timing, etc.,
  specialized for each benchmark, not as part of the rules for the
  application logic.}

In Alda, such rules and tricks are never needed.  A call to \co{\INFER}
with query \co{query9\_1} returns the set of all query results as desired.
If \co{query9\_1} is a non-local predicate, then the set value of
\co{query9\_1} can be used directly, and no explicit call to \co{\INFER} is
needed.
\arxiv{In case only one result is wanted, a special function for taking any
  one value can be applied to the result set of calling \co{\INFER} or the
  non-local predicate; an optimized implementation can then search for only
  the first result.}

\mypar{Function symbols}
Logic rules may use function symbols to form structured data
that
can be used as
arguments to predicates.  Uses of function symbols can be translated away.
The Mondial benchmark uses a function symbol \co{prov} in several intermediate conclusions
and hypotheses of the form \co{isa(prov(Y,X),provi)} or \co{att(prov(Y,X),number,A)}.
They can simply be translated to 
\co{isa('prov',Y,X,provi)} and \co{att('prov',Y,X,number,A)}, respectively.

\mypar{Negation}
\arxiv{Logic languages may use negation applied to hypotheses in rules. }%
Most logic languages only support non-stratified negation, where there is
no negation involved in cyclic dependencies among predicates.  Such
negation can be done by set differences.  The ModSG benchmark has such a
negation, as follows, where \co{sg} is defined by the rules in the SG
benchmark, and \co{nonsg} is defined by two new rules:
\begin{code}
  sg2(X,Y) :- sg(X,Y), not nonsg(X,Y).
\end{code}
In Alda, this can be written as
\begin{code}
  sg2 = \INFER(sg, \RULES=sg\_rs) - \INFER(nonsg, \RULES=nonsg\_rs)
\end{code}
where \co{sg\_rs} and \co{nonsg\_rs} are the rule sets
defining \co{sg} and \co{nonsg}, respectively, and all base predicates are
non-local.

\arxiv{We also added support for negation in our implementation, which
  translates negation to tabled negation \co{tnot} in XSB, instead of
  Prolog's negation as failure.  This handles even non-stratified negation
  by computing well-founded semantics using XSB.  The Win and MagicSet
  benchmarks have non-stratified negation.
  Both of them, as well as ModSG, can be expressed directly in Alda
  rule sets by using \co{not} for negation.
}

\mypar{Benchmarking and organization} %
In OpenRuleBench benchmarks, even though the rules to be benchmarked are
declarative and succinct, the benchmarking code for reading input, running
tests, timing, and writing results are generally much larger.  For example,
the Join1 benchmark has 4 small rules and 9 small queries similar in size
to those in the transitive closure example, plus a manually added tabling
directive for optimization.  However, for each query, 19 more lines for an
import and two much larger rules are used to do the reading, running,
timing, and writing.

In general, because benchmarking executes a bundle of commands, 
scripting those directly is simplest.  
Furthermore, organizing benchmarking code using procedures, objects, etc., 
allows easy reuse without duplicated code.  
These features are much better supported, for both ease of programming 
and performance, in languages like Python than rule systems.

In fact, OpenRuleBench uses a large number of many different files, 
in several languages (language of the system being tested, XSB, shell script, 
Python, makefile) for such scripting.  
For example for Join1, the 4 rules, tabling directive, and benchmarking code
are also duplicated in each of the 9 XSB files, one for each query; 
a 46-line shell script and a 9-line makefile are also used.

In contrast, our benchmarking code is in Alda, which uses Python functions
for scripting.\oops{}
A same 45-line Alda program is used %
for timing any of the benchmarks, and %
for pickling (i.e., object serialization in Python, for fast data 
reading after the first reading) and timing of pickling.

\mypar{Aggregation}
Despite the wide variety of benchmarks in OpenRuleBench, 
no benchmark uses aggregate queries.  Aggregate queries are essential for
many database, data mining, and machine learning applications.  We discuss
them and compare with aggregate queries in a rule language like XSB in
\arxiv{Section~\ref{sec-pa}}\oops{}.

\arxiv{
\oops{\section{Additional programming and performance benchmarks}}

\mysubsec{RBAC---rules with objects, updates, and set queries}
\label{sec-rbac}

As discussed in Section~\ref{sec-obj}, a complex and inefficient
\co{\WHILE} loop was used in \cite{LiuSto07RBAC-ONR} to program the
transitive role hierarchy, but as discussed in Section~\ref{sec-prob}, an
efficient algorithm with appropriate data structures and updates would be
drastically even more complex.

With support for rules, we easily wrote the entire RBAC standard in Alda,
similar as in Python~\cite{LiuSto07RBAC-ONR}, except with rules for
computing the transitive role hierarchy, as described in
Section~\ref{sec-obj},\arxiv{ and with simpler set queries and omission of
  \co{self},} despite complex class inheritance relationships, yielding a
simpler yet more efficient program.

Below, we describe different ways of using rules to compute the transitive
role hierarchy and the function\oops{\linebreak[5]}
\co{AuthorizedUsers(role)} in Section~\ref{sec-obj}.
All these ways are declarative and differ in size by only 1-2 lines.
Table~\ref{tab-rbac} summarizes the benchmarks for RBAC that include all
RBAC classes with their inheritance relationships and perform update
operations and these query functions in different ways.

In particular, in the first way below, a field, \co{transRH}, is used and
maintained automatically; it avoids calling \co{transRH()} repeatedly, as
desired in the RBAC standard, and it does so without the extra maintenance
code in the RBAC standard for handling updates.

\begin{table}[t]
  \small
  \centering
\begin{tabular}{@{~}l|@{~}p{\arxiv{69}\oops{40}ex}}
  Name    & Features used for computing transitive role hierarchy
  \\\hline
  RBACnonloc & rule set \co{transRH\_rs} with %
             implicit \co{\INFER}, without \co{transRH()}, Sec.\,\ref{sec-rbac}
  \\\hline
  RBACallloc   & rule set \co{trans\_role\_rs} %
             and \co{transRH()} that has only \co{\INFER}, Sec.\,\ref{sec-rbac}
  \\\hline
  RBACunion & rule set \co{trans\_rs} and 
             \co{transRH()} that has \co{\INFER} and union, Sec.\,\ref{sec-obj}
  \\\hline
  RBACda    & \co{\WHILE} loop and 
             high-level set queries in DistAlgo, Sec.\,\ref{sec-pred}
  \\\hline
  RBACpy    & \co{\WHILE} loop and
             high-level set comprehensions in Python~\cite{LiuSto07RBAC-ONR}
  \\\hline
\end{tabular}
\caption{Benchmarks for RBAC updates and queries.  
  Each performs a combination of updates to sets and relations \co{USERS},
  \co{ROLES}, \co{UR}, and \co{RH} and queries with function
  \co{AuthorizedUsers(role)}, where the transitive role hierarchy is computed
  with a different way of using rules, or not using rules.
  In \co{AuthorizedUsers(role)} of all five programs, the call to
  \co{transRH()}, or reference to field \co{transRH}, is lifted out of the
  set query, by assigning its value to a local variable and using that
  variable in the query.}
\label{tab-rbac}
\end{table}

\mypar{Rules with only non-local predicates}
\label{sec-nonlocal}
Using rules with only non-local predicates, one can use a relation
\co{transRH} in place of calls to \co{transRH()}, e.g., in function
\co{AuthorizedUsers(role)},
by simply adding a field \co{transRH} and using the following rule set in
class \co{HierRBAC}:\footnote{The first rule could actually be omitted,
  because
  the second argument of \co{RH} is always in \co{ROLES} and thus the
  second rule when joining \co{RH} with reflexive pairs in \co{transRH}
  from the third rule subsumes the first rule.}
\begin{lstlisting}
  rules transRH_rs:  # no need to use infer explicitly
    transRH(x,y) if RH(x,y)
    transRH(x,y) if RH(x,z), transRH(z,y)
    transRH(x,x) if ROLES(x)
\end{lstlisting}
Field \co{transRH} is automatically maintained at updates to \co{RH} and
\co{ROLES}
by implicit calls to \co{\INFER}; %
no explicit calls to \co{\INFER} are needed.  
This eliminates the need of function \co{transRH()} and repeated expensive
calls to it even when its result is not changed most of the time.
Overall, this simplifies the program, ensures correctness, and improves
efficiency.

\mypar{Rules with only local predicates}
\label{sec-appl-local}
Using rules with only local predicates, \co{\INFER} must be called
explicitly.
One can simply use the function \co{transRH()} in Section~\ref{sec-obj},
which calls \co{\INFER} using rule set \co{trans\_rs} in the running example
and then unions with reflexive role pairs.  Alternatively, one can use the
rules in \co{trans\_rs} plus a rule that uses a local \co{role} set
that adds the reflexive role pairs:
\begin{oopsenv}
\begin{lstlisting}
rules trans_role_rs: # as trans_rs plus the added last rule
  path(x,y) if edge(x,y)
  path(x,y) if edge(x,z), path(z,y)
  path(x,x) if role(x)               
def transRH():        # use infer only, pass in also ROLES
  return infer(path, edge=RH, role=ROLES, rules=trans_role_rs)
\end{lstlisting}
\end{oopsenv}
\begin{arxivenv}
\begin{lstlisting}
  rules trans_role_rs:  # as trans_rs plus the added last rule
    path(x,y) if edge(x,y)
    path(x,y) if edge(x,z), path(z,y)
    path(x,x) if role(x)               
  def transRH():        # use infer only, pass in also ROLES
    return infer(path, edge=RH, role=ROLES, rules=trans_role_rs)
\end{lstlisting}
\end{arxivenv}
Both ways show the ease of using rules by simply calling \co{\INFER}.
Despite possible inefficiency in some uses, %
using only local predicates has the advantage of full reusability of rules 
and full control of calls to \co{\INFER}.

\mypar{Rules with both local and non-local predicates}
\label{sec-appl-mixed}
One can also use rules with a combination of local and non-local
predicates, e.g., the same rules as above but with field \co{ROLES} in
place of the local \co{role},
removing the need for \co{\INFER} to pass in \co{ROLES}.
Any other combination %
can also be used.
Different combinations give different controls to \co{\INFER} to pass in and
out appropriate sets. %

Of course,
non-recursive set queries, such as\oops{\linebreak[5]} \co{AuthorizedUsers(role)} can also be
expressed using rules, and use any combination of local and non-local
predicates.

\mysubsec{Program analysis---rules with aggregate queries and recursive
  functions}
\label{sec-pa}

We have used Alda to analyze widely-used Python packages, %
and designed a benchmark %
based on our experiences, especially to show
integrated use of rules with aggregate queries and recursive
functions.
Aggregate queries help quantify and characterize the analysis results,
and recursive functions help do these on recursive structures.

The benchmark is for analysis of class hierarchy of Python programs.  It
uses logic rules to extract class names and construct the class extension
relation;
aggregate queries and set queries to characterize the results and find
special cases of interest;
recursive functions as well as aggregate and set queries to analyze the
special cases;
and more logic rules, functions, and set and aggregate queries to further
analyze the special cases.

Table~\ref{tab-PA} summarizes different parts of this benchmark, called PA.
We also use a variant, called PAopt, that is the same as PA except that, in
the recursive rule for defining transitive descendant relationship, the two
hypotheses are reversed, following previously studied
optimizations~\cite{LiuSto09Rules-TOPLAS,TekLiu10RuleQuery-PPDP}.
Additionally, we compare with corresponding programs written in a rule
language like XSB, expressing aggregate queries and reursive functions.
\begin{table}[t]
  \small
  \centering
\begin{tabular}{@{~}l@{}@{~}l|@{~}p{\arxiv{32}\oops{18}ex}@{~}|@{~}p{\arxiv{52}\oops{24}ex}@{~}}
  \Hex{0}Part\Hex{-4}  & & Analysis & Features used \\\hline
  1 & Ext  & classes,\oops{\newline} extension relation 
                  & rules (recursive if refined name analysis is used)
  \\\hline
  2 & Stat & statistics, roots
                  & aggregate and set queries
  \\\hline
  3 & Hgt & max height,\oops{\newline} roots of max height 
                  & recursive functions,\oops{\newline} aggregate and set queries
  \\\hline
  4 & Desc & max desc,\oops{\newline} roots of max desc
                  & recursive rules, functions,\oops{\newline} aggregate and set queries
  \\\hline
\end{tabular}
\caption{Benchmark PA for program analysis, integrating 
  different kinds of analysis problems.
  In 1 and 4 that use rules,
  not using rules  (esp.\ for recursive analysis, with tabling) 
  would be drastically worse
  (i.e., harder to program and less efficiency).
  In 2-4 that use aggregate and set queries, using
  rules or recursive functions would be clearly worse.
  In 3 and 4 that use functions, not using functions (with tabling, also
  called caching) would be much worse.}
\label{tab-PA}
\end{table}

Because the focus is on evaluating the integrated use of different
features, each part that uses a single feature, such as rules, is
designed to be small.  Compared with making each part larger, which
exercises individual features more, this design highlights %
the overhead of
connecting different parts, in terms of both ease of use and efficiency of
execution. %

The program takes as input %
the abstract syntax tree (AST) of a Python program 
(a module or an entire package), represented as set of facts.
Each AST node of type \p{T} with \p{k} children corresponds to 
a fact for predicate \p{T}
with \p{k}+1 arguments: id of the node, and
ids the \p{k} children. %
Lists are represented using \co{Member} facts.  A
\co{Member(\p{lst},\p{elem},\p{i})} fact denotes that list \p{lst} has
element \p{elem} at the \p{i}th position.

\mypar{%
Part 1: Classes and class extension relation} 
This examines all \co{ClassDef} nodes in the AST.  A \co{ClassDef} node has
5 children: class name, list of base-class expressions, and three nodes not
used for this analysis.
The following rules can be used to find all defined class names and build
a class extension relation using base-class expressions that are
\co{Name} nodes.  A \co{Name} node has two children: name and
context.
\begin{lstlisting}
  rules class_extends_rs:
    defined(c) if ClassDef(_, c,_, _,_,_)
    extending(c,b) if ClassDef(_, c,baselist, _,_,_),
                      Member(baselist,base,_), Name(base,b, _)
\end{lstlisting}
For a dynamic language like Python, analysis involving names can be refined
in many ways to give more precise results,
e.g.,~\cite{Gor+10Alias-DLS}.
We do not do those here, but Datalog rules are particularly good for
such analysis of bindings and aliases for names,
e.g.,~\cite{smara15pointer}.

\mypar{%
Part 2: Characterizing results and finding special cases}
This computes statistics for defined classes and the class extension
relation
and finds %
root classes (class with subclass but not super class).
These use aggregate queries and set queries.
\begin{lstlisting}
  num_defined := count(defined)
  num_extending := count(extending)
  avg_extending := num_extending/num_defined
  roots := {c: (_,c) in extending, not some (=c,_) in extending}
\end{lstlisting}
Similar queries can compute many other statistics and cases: maximum number
of classes that any class extends, leaf classes, histograms, etc.

\mypar{Part 3: Analysis of special cases}
This computes the maximum height of the extension relation, which is the
maximum height of the root classes, and finds root classes of the maximum
height.
These use a recursive function as well as aggregate and set queries.
\begin{lstlisting}
  def height(c):
    return 0 if not some (_,=c) in extending
           else 1 + max{height(d): (d,=c) in extending}

  max_height := max{height(r): r in roots}
  roots_max_height := {r: r in roots, height(r) = max_height}
\end{lstlisting}
For efficiency, when a subclass can extend multiple base classes, caching of
results of function calls is used.
In Python, one can simply add \co{import functools} to import module
\co{functools}, and add \co{@functools.cache} just above the definition of
\co{height} to %
cache the results of that function.

\mypar{Part 4: Further analysis of special cases}
This computes the maximum number of descendant classes following the
extension relation from a root class, and finds root classes of the maximum
number.
Recursive functions and aggregate queries similar to finding maximum height
do not suffice here,
due to shared subclasses that may be at any depth.
Instead, the following rules can infer all \co{desc(c,r)} facts where class
\co{c} is a descendant following the extension relation from root class
\co{r}, and aggregate and set queries with function \co{num\_desc} then
compute the desired results.
\begin{lstlisting}
  rules desc_rs:
    desc(c,r) if roots(r), extending(c,r)
    desc(c,r) if desc(b,r), extending(c,b) 
\end{lstlisting}
\begin{lstlisting}
  def num_desc(r):
    return count{c: (c,=r) in desc}

  max_desc := max{num_desc(r): r in roots}
  roots_max_desc := {r: r in roots, num_desc(r) = max_desc}
\end{lstlisting}
For efficiency of the last query, caching is also used for function
\co{num\_desc}.  If the last query is omitted, function
\co{num\_desc} can also be inlined in the \co{max\_desc} query.

\arxiv{
Benchmark PAopt is the same as PA except that in rule set \co{desc\_rs},
the two hypotheses in the second rule are reversed; this allows default
indexing in XSB, which is on the first argument, to find matching values
\co{c}, \co{b}, \co{r} faster in that order.}

\mypar{Comparing with aggregate queries and functions in rule languages}
\arxiv{While rules in Alda correspond directly to rules in rule languages,
  expressing aggregate queries and functions using rules require
  translations that formulate computations as hypotheses and introduce
  additional variables to relate these hypotheses.}

Aggregate queries are used extensively in database %
and machine learning applications, and are essential for analyzing large data
or uncertain information.  
While these queries are easy to express directly in database languages and
scripting languages, they are less so in rule languages like Prolog; most
rule languages also do not support general aggregation with recursion due
to their subtle semantics~\cite{LiuSto22RuleAgg-LFCS}.  For example, the
simple query \co{num\_defined = count(defined)} in Alda, if written in XSB,
would become:
\begin{code}
  num_defined(N) :- setof(C, defined(C), S), length(S, N).
\end{code}

Recursive functions are used extensively in list and tree processing and in
solving divide-and-conquer problems.  They are natural for computing
certain information about parse trees, nested scopes,
etc.
However, in rule languages, they are expressed in a way that mixes function
arguments and return values, and require sophisticated mode analysis to
differentiate arguments from returns.  For example, the \co{height} query,
if written in XSB, would become:
\begin{code}
  height(C,0) :- not(extending(_,C)).
  height(C,H) :- findall(H1, (extending(D,C), height(D,H1)), L),
                 max_list(L,H2), H is H2+1.
\end{code}\Vex{-1}

}

\notes{}%

\notes{} %

\section{Experimental evaluation} %
\label{sec-expe}

We have implemented a prototype compiler for Alda.
It generates executable code in Python.  The generated code calls the XSB
logic rule engine~\cite{swift2012xsb,xsb21} for inference using rules.

We implemented Alda by extending the DistAlgo
compiler~\cite{Liu+12DistPL-OOPSLA,Liu+17DistPL-TOPLAS,distalgo22git}.
DistAlgo is an extension of Python with high-level set queries as well as
distributed processes.  The %
compiler is implemented in Python~3, and uses the Python parser.  
So Python syntax is used in place of the ideal syntax presented 
in Section~\ref{sec-lang}. 

The Alda implementation extends the DistAlgo compiler to support rule-set
definitions, function \co{\INFER}, and updates to non-local variables.
It handles only direct updates to variables used as predicates, 
not updates through aliasing; a previous
alias analysis~\cite{Gor+10Alias-DLS} that took several years to develop
was only for Python 2.
Currently\oops{} Datalog rules\arxiv{ extended with negation} are supported,
but extensions for\arxiv{ more} general 
rules can be handled similarly,
and inference using XSB can remain the same.
Calls to \co{\INFER} are automatically added at updates to non-local base
predicates of a rule set.

In particular, the following Python syntax is used for rule sets, where a
rule can be either one of the two forms below, so the only restriction is
that the name \co{rules} %
is reserved.\Vex{-1}

\begin{code}
  def rules (name = \p{rsname}):
    \p{conclusion}, if_(\p{hypothesis\sb{1}}, \p{hypothesis\sb{2}}, \p{...}, \p{hypothesis\sb{h}})
    if (\p{hypothesis\sb{1}}, \p{hypothesis\sb{2}}, \p{...}, \p{hypothesis\sb{h}}): \p{conclusion}
    \p{...}
\end{code}\Vex{-1}
Rule sets are translated into Prolog rules at compile time.  
The compiler directive \co{:-~auto\_table.} is added for automatic tabling
in XSB.

For function \co{\INFER}, the implementation translates the values of
predicates and the list of queries into facts and queries in standard
Prolog syntax, and translates the query answers back to values of set
variables.  It invokes XSB using a command line in between, passing data
through files; this external interface has an obvious overhead, but it has
not affected Alda having generally good performance.  \co{\INFER}
automatically reads and writes non-local predicates used in a rule set.

Note that the overhead of the external interface can be removed with an
in-memory interface from Python to XSB, which is actively being developed
by the XSB team\footnote{A version working for Unix, not yet Windows, has
  just been released, and passing data of size 100 million in memory took
  about 30 nanoseconds per element~\cite[release notes]{xsb22}.  So even
  for the largest data in our experiments, of size a few millions, it would
  take 0.1-0.2 seconds to pass in memory, instead of 10-20 seconds with the
  current external interface.}.
However, even with the overhead of the external interface, Alda is still
faster or even drastically faster than half or more of the rule engines
tested in OpenRuleBench~\cite{Lia+09open} for all benchmarks measured
except DBLP (even though OpenRuleBench uses the fastest manually optimized
program for each problem for each rule engine), and than not using rules at
all (without manually writing or adapting a drastically more complex,
specialized algorithm implementation for each problem).

\notes{}

Building on top of DistAlgo and XSB, the compiler consists of about 
1100
lines of Python and about 50 lines of XSB.
This is owing critically to the
overall framework and comprehensive functions, especially support for
high-level queries, already in the DistAlgo compiler and to the powerful
search and inference engine of XSB.  The parser for the rule extension is
about 270 %
lines, and 
update analysis and code generation for rules and inference are about 800
lines.

The current compiler does not perform further optimizations, because they
are orthogonal to the focus of this paper, and our experiments already
showed generally good performance.  Further optimizations can be
implemented in either the Alda compiler to generate optimized rules
and tabling and indexing directives, or in XSB.  Incremental maintenance
under updates can also be implemented in either one, with a richer
interface between the two.

We discuss our experiments on the benchmarks described in
Section~\ref{sec-bench}.
The experiments selected are meant to show acceptable performance even
under the most extreme overhead penalties we have encountered.  Our
extensive experiments with other uses of Alda
have experienced minimum such penalties.

All measurements were taken on\arxiv{
a machine with an Intel Xeon X5690 3.47 GHz CPU, 94 GB RAM,
running 64-bit Ubuntu 16.04.7}\oops{},
Python 3.9.9, and XSB 4.0.0.
For each experiment, the reported running times are CPU times averaged over
\oops{}\arxiv{10} runs.  Garbage collection in Python was disabled for
smoother running times.
Program sizes are numbers of lines excluding comments and empty lines.
Data sizes are number of facts.\notes{}

\mysubsec{Compilation times and program sizes}
\begin{table}[tp]
  \arxiv{\small}\oops{}
  \centering
\begin{tabular}{@{~}l|r||r|r|r|r}
  \arxiv{\Hex{1}}Benchmark
             & Original~ & Alda  & Compilation & Generated~ & Generated\\
  \Hex{\arxiv{4}\oops{}}name 
             & XSB size\,& size\,& time (ms)~\,& Python size & XSB size\,\\
  \hline\hline
  Join1      &  225 &  23 &  33.037 &  32 &   5\\\hline
  Join2      &   41 &  11 &  18.540 &  16 &   9\\\hline
  JoinDup    &  163 &  42 &  45.580 &  20 &  36\\\hline  
  LUBM       &  377 & 125 & 116.378 &  29 & 110\\\hline
  Mondial    &   36 &   8 &  16.225 &  16 &   6\\\hline
  DBLP       &   20 &   4 &  16.319 &  16 &   2\\\hline %
  TC         &   75 &   5 &   7.920 &  16 &   3\\\hline
  TCrev      &  *75 &   5 &   7.712 &  16 &   3\\\hline
  TCda       &   -- &  23 &   4.851 &  47 &   -\\\hline
  TCpy       &   -- &  25 &   6.506 &  39 &   -\\\hline %
  SG         &   90 &  13 &  17.939 &  20 &   7\\\hline
  WordNet    &  298 &  58 &  76.226 &  44 &  28\\\hline  
  Wine       & 1103 & 970 & 605.613 &  16 & 968\\\hline %
  ModSG      &   38 &  14 &  14.779 &  16 &  12\\\hline %
  Win        &   24 &   4 &   9.477 &  16 &   2\\\hline
  MagicSet   &   34 &   9 &  18.210 &  16 &   7\\\oops{}\arxiv{\hline
  ORBtimer   &   -- &  45 &  35.178 &  56 &   -\\\hline %
  \hline                               %
  RBACnonloc &   -- & 423 & 346.377 & 538 &   4\\\hline
  RBACallloc &   -- & 387 & 318.403 & 481 &   4\\\hline
  RBACunion  &   -- & 386 & 316.557 & 481 &   3\\\hline
  RBACda     &   -- & 385 & 312.787 & 483 &   -\\\hline
  RBACpy     &   -- & 387 & 314.561 & 476 &   -\\\hline
  RBACtimer  &   -- &  44 &  43.258 &  67 &   -\\\hline %
  \hline                   
  PA         &  *55 &  33 &  49.695 &  93 &   6\\\hline %
  PAopt      &  *55 &  33 &  40.848 &  93 &   6\\\hline %
  PAtimer    &   -- &  40 &  32.624 &  56 &   -\\\hline %
  \hline
}
  \end{tabular}
  \caption{Compilation times and program sizes before and after compilation. 
    For Original XSB size, entries without * are from OpenRuleBench,
    as in Table~\ref{tab-rulebench};
    * indicates XSB programs we added;
    -- means there is no corresponding XSB program.
    For Generated XSB size, - means no XSB code is generated.
    \oops{}}
  \label{tab-compile}
\end{table}

Table~\ref{tab-compile} shows the compilation times and program sizes
before and after compilation, for all three sets of benchmarks described in
Section~\ref{sec-bench}, plus three variants of TC in the first set as
explained in Section~\ref{sec-expe-tc}.
\arxiv{For each set of benchmarks, there is a shared file of benchmarking
code,
shown in the last row of each set;
for OpenRuleBench, ORBtimer includes 17 lines for pickling and timing of
pickling.}

The compilation times for all benchmark programs are 0.6 seconds or less,
and for all but Wine and RBAC benchmarks about 0.1 seconds or less.
\oops{}

\arxiv{
For OpenRuleBench benchmarks, Alda program sizes are all smaller than
the corresponding XSB sizes, almost all by dozens or even hundreds of lines, 
and by an order of magnitude for Join1 and TC.  In place of the extra XSB
code for benchmarking and manually added optimization directives, all Alda
programs use the single shared 45-line file, ORBtimer, for benchmarking
and pickling.
The generated XSB size is exactly the number of rules plus one line for
\co{:- auto\_table.}, for each rule set.
The generated Python size for benchmarks from OpenRuleBench is larger for benchmarks with more queries.  %

For RBAC benchmarks, all Alda sizes include 3 files of 373 lines total for
all 9 RBAC classes, taking a total compilation time of 299.503
milliseconds,
generating a total compiled Python size of 456 lines.
Each way of computing the query functions is in a separate class extending
Hierarchical RBAC; RBACnonloc is over 30 lines more than others because all
query functions in Hierarchical RBAC, not just \co{authorizedUsers(role)},
are overridden to use field \co{transRH} in place of calls to
\co{transRH()}.

For PA benchmarks, the benchmarking code for the XSB programs is written in
a similar way as the benchmarks in OpenRuleBench, and takes 23 lines for
each benchmark.
}

\mysubsec{Performance of classical queries using rules}
\label{sec-expe-tc}
To evaluate the efficiency of classical queries using rules in Alda, we use
four programs for computing transitive closure: (1) TC---the TC benchmark
from OpenRuleBench, which is the same as \co{trans\_rs} except with renamed
predicates, (2) TCrev---same as \co{trans\_rs} but reversing the two
predicate names
in the recursive rule,\arxiv{
a well-known variant,}
(3) TCda---\co{\WHILE} loops with high-level queries
in DistAlgo as in Section~\ref{sec-pred}, and (4) TCpy---\co{\WHILE} loops
with comprehensions in Python as in Section~\ref{sec-prob}.  

For comparison, we also directly run the XSB program for TC from
OpenRuleBench, and its corresponding version for TCrev, except we change
\co{load\_dyn} used in OpenRuleBench to \co{load\_dync}, for much faster
reading of facts in XSB's canonical form; we call these two programs TCXSB
and TCrevXSB, respectively.  Note that XSB programs in OpenRuleBench, not
using \co{load\_dync}, are significantly slower for large input, e.g., see
the DBLP benchmark in Section~\ref{sec-expe-scale}.

We use the data generator in OpenRuleBench to generate data\oops{}.%
\arxiv{
The generator is sophisticated in trying to ensure randomness 
as well as cyclic vs.\ acyclic cases.
We use the same number of vertices, 1000, and a range of numbers of edges, 
10K to 100K, %
to include the first of two data points (50K and 500K edges)
reported in~\cite{Lia+09open}. 
For the cyclic graphs generated, even for the smallest data of 10K edges,
i.e., each vertex having edges going to only 1\% of vertices---10 out of
1000---on average, the resulting transitive closure is already the complete
graph of 1M edges.
}

Because of interfacing with XSB through files, the total running time of
Alda programs includes not only (1) reading data, (2) executing queries,
and (3) returning results, but also (2pre) preparing data, queries, and
commands and writing data
to files for XSB to start and read before (2),
and (2post) reading results from files written by XSB after (2).
We report the total running time as well as separate times for pickling
and for interfacing with XSB.

\begin{figure}[t]
  \small
  \center
\newcommand\figTC[1]{
\begin{tikzpicture}[every mark/.append style={mark size=2pt}]
  \begin{axis}[
    ymin=0, ymax=32,
    xtick={10,20,30,40,50,60,70,80,90,100},
    xlabel={Number of edges (in thousands)},
    ylabel={CPU time (in seconds)},
    ylabel shift=-1ex,
    tick label style={font=\arxiv{\scriptsize}\oops{}},
    legend pos=north west,
    legend cell align={left},
    legend style={font=\arxiv{\scriptsize}\oops{}, row sep=-.5ex},
    ymajorgrids=true, %
    ]
\addplot[mark=pentagon,color=red] table[x=edges,y=TC,col sep=comma] {#1};
\addplot[mark=star,color=ForestGreen,densely dashed] table[x=edges,y=TCrev,col sep=comma] {#1};
\addplot[mark=x,color=ForestGreen,densely dashed] table[x=edges,y=TCrevXSB,col sep=comma] {#1};
\addplot[mark=diamond,color=red] table[x=edges,y=TCXSB,col sep=comma] {#1};
\addplot[mark=square,color=red] table[x=edges,y=TC_extra,col sep=comma] {#1};
\addplot[mark=+,color=ForestGreen,densely dashed] table[x=edges,y=TCrev_extra,col sep=comma] {#1};
\addplot[mark=triangle] table[x=edges,y=rawR,col sep=comma] {#1};
\addplot[mark=o] table[x=edges,y=pickleW,col sep=comma] {#1};
\legend{TC, TCrev, TCrevXSB, TCXSB, TC\_extra, TCrev\_extra\!, rawR, pickleW}
\end{axis}
\end{tikzpicture}\Vex{-1}\oops{}
}
\figTC{TCcyc.csv}
\figTC{TCacyc.csv}
\caption{Running times of TC benchmarks on cyclic (left) and acyclic
  (right) graphs.  TCpy and TCda are not shown because they are
  asymptotically slower and took drastically longer: on 100 vertices, 
  with cyclic data of 200 edges (2\% of smallest data point shown),
  TCpy took 624.6 seconds, and TCda took 249.9 seconds; and with
  acyclic data of 600 edges, %
  TCpy took 160.4 seconds, and TCda took 65.2 seconds.}
\label{fig-TC}
\end{figure}
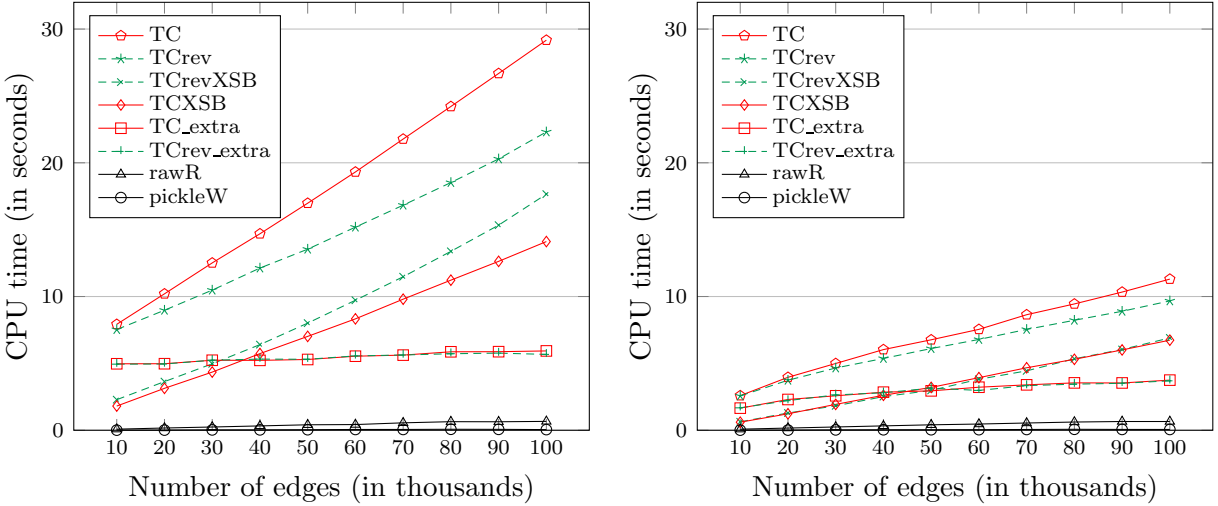

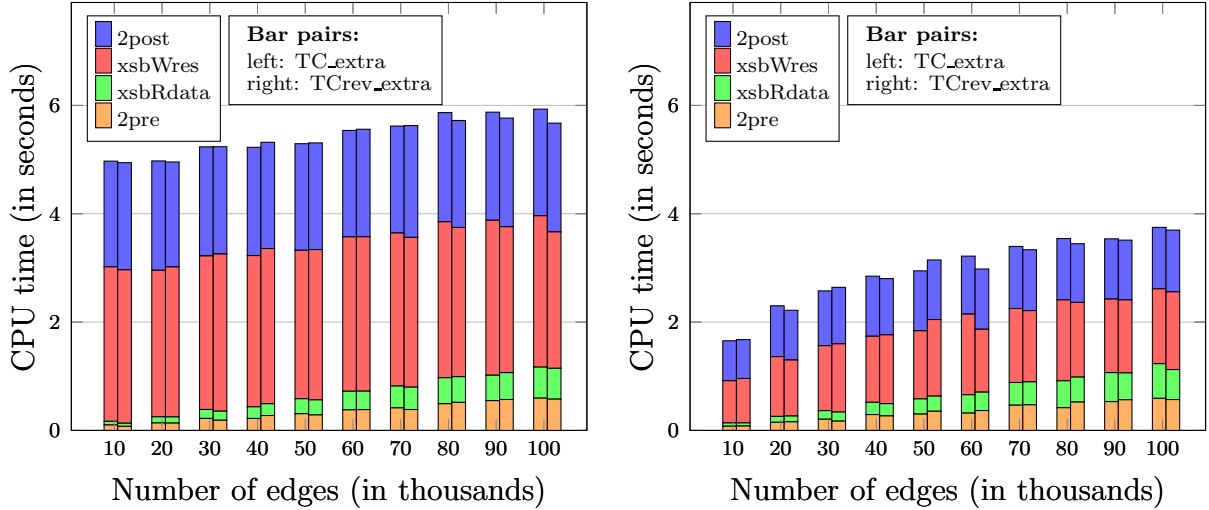
\begin{figure}[ht]
  \centering
\newcommand\figTCex[1]{
\begin{tikzpicture}[
  every axis/.style={
    ybar stacked, bar width=1ex,
    ymin=0, ymax=7.9,
    xtick=data, xticklabels from table={#1}{edges},
    xlabel={Number of edges (in thousands)},
    ylabel={CPU time (in seconds)},
    ylabel shift=-1ex,
    tick label style={font=\arxiv{\scriptsize}\oops{}},
    legend style={cells={anchor=west}, legend pos=north west},
    legend style={font=\arxiv{\scriptsize}\oops{}, row sep=-.5ex},
    reverse legend=true,
  }
  ]
\begin{axis}[bar shift=-.25ex,
  ymajorgrids=true, %
  ]
  \addplot[fill=orange!60] table[y=2pre,meta=edges,x expr=\coordindex] {#1};
  \addplot[fill=green!60] table[y=xsbRdata,meta=edges,x expr=\coordindex]{#1};
  \addplot[fill=red!60] table[y=xsbWres,meta=edges,x expr=\coordindex] {#1};
  \addplot[fill=blue!60] table[y=2post,meta=edges,x expr=\coordindex] {#1};

  \legend{2pre, xsbRdata\!, xsbWres, 2post}
  \node[draw,fill=white,anchor=north east] at (rel axis cs: 0.72,0.97){\shortstack[l]{
      {\arxiv{\scriptsize}\oops{} {\bf Bar pairs:}}\\
      {\arxiv{\scriptsize}\oops{} left: TC\_extra}\\
      {\arxiv{\scriptsize}\oops{} right: TCrev\_extra}
    }};
\end{axis}
\begin{axis}[bar shift=.75ex]
  \addplot[fill=orange!60] table[y=rev2pre,meta=edges,x expr=\coordindex] {#1};
  \addplot[fill=green!60] table[y=revxsbRdata,meta=edges,x expr=\coordindex]{#1};
  \addplot[fill=red!60] table[y=revxsbWres,meta=edges,x expr=\coordindex] {#1};
  \addplot[fill=blue!60] table[y=rev2post,meta=edges,x expr=\coordindex] {#1};
\end{axis}
\end{tikzpicture}\Vex{-1}
}
\figTCex{\TCexCyc}
\figTCex{\TCexAcyc}
\caption{Breakdown of TC\_extra and TCrev\_extra.}
\label{fig-TC-break}
\end{figure}

Figure~\ref{fig-TC} shows the running times of the TC benchamrks. %
\oops{}\arxiv{%
RawR and PickleW are times for reading facts in XSB/Prolog form as used
in OpenRuleBench and writing them in pickled form for use in Alda,
respectively. Pickling is done only once; the pickled data is read in all
repeated runs and all of TC, TCrev, TCda, and TCpy.
TC\_extra and TCrev\_extra are the part of TC and TCrev, respectively, for
extra work interfacing with XSB, i.e., for 2pre and 2post and for XSB to
read data (xsbRdata) and write results (xsbWres).
Figure~\ref{fig-TC-break} shows the breakdown of TC\_extra and TCrev\_extra
among 2pre, 2post, xsbRdata, and xsbWres.
}

\oops{}

\oops{}

\pgfplotstableread{
edges 2pre 2post xsbRdata xsbWres rev2pre rev2post revxsbRdata revxsbWres
 10  0.078 0.737  0.060 0.780   0.082 0.718  0.057 0.820
 20  0.150 0.938  0.109 1.104   0.159 0.918  0.107 1.034
 30  0.207 1.012  0.158 1.199   0.172 1.043  0.168 1.259
 40  0.293 1.109  0.226 1.220   0.268 1.038  0.226 1.271
 50  0.302 1.106  0.280 1.258   0.354 1.100  0.280 1.413
 60  0.320 1.071  0.339 1.488   0.367 1.112  0.340 1.161
 70  0.466 1.145  0.419 1.366   0.473 1.124  0.424 1.315
 80  0.419 1.133  0.498 1.493   0.524 1.083  0.461 1.379
 90  0.532 1.112  0.535 1.358   0.566 1.103  0.496 1.348
100  0.592 1.135  0.640 1.382   0.570 1.139  0.554 1.435
}\TCexAcyc
\pgfplotstableread{
edges 2pre 2post xsbRdata xsbWres rev2pre rev2post revxsbRdata revxsbWres
 10  0.101 1.951  0.067 2.852   0.076 1.978  0.057 2.832
 20  0.138 2.018  0.113 2.706   0.136 1.938  0.114 2.768
 30  0.222 2.013  0.165 2.835   0.188 1.980  0.168 2.902
 40  0.221 1.999  0.215 2.792   0.275 1.962  0.218 2.865
 50  0.308 1.967  0.277 2.741   0.286 1.969  0.278 2.775
 60  0.380 1.964  0.343 2.851   0.384 1.984  0.341 2.851
 70  0.416 1.974  0.404 2.825   0.386 2.064  0.413 2.765
 80  0.493 2.015  0.480 2.879   0.518 1.975  0.474 2.755
 90  0.550 1.995  0.469 2.861   0.571 2.005  0.496 2.694
100  0.596 1.969  0.573 2.794   0.578 2.008  0.569 2.518
}\TCexCyc

\arxiv{
Not shown in Figure~\ref{fig-TC}, but the times in TC and TCrev for reading
facts are similar to PickleW; 
the times in TCXSB and TCrevXSB for reading facts are similar to RawR.
The remaining times in TCXSB and TCrevXSB are XSB query times
only, because XSB programs in OpenRuleBench do not output or even collect
the query result in any way.

Also not shown in Figures~\ref{fig-TC} and~\ref{fig-TC-break} are the times
in TC and TCrev for starting the XSB process and, as part of 2pre, for
preparing the queries to start XSB with proper arguments and status checks.
These are small, at 0.1--0.2 seconds and 0.03--0.04 seconds, respectively,
because XSB is invoked only once in each run.}

We observe that:
\begin{itemize}
\item \oops{}\arxiv{%
The extra times interfacing with XSB are obvious, %
here dominated by passing query results by files (xsbWres and 2post) 
as shown in Figure~\ref{fig-TC-break}
because the results of transitive closure are generally large, 
and larger for cyclic graphs}.
This overhead of going through files will be removed when using direct
mapping between XSB and Python data structures in memory.

\item The remaining times without the interfacing overhead are basically
  all XSB query times for the different XSB programs.
In particular, 
TCrev is faster than TC in Alda, %
but TCXSB is faster than TCrevXSB.
This is because OpenRuleBench %
uses the fastest manually optimized program for each problem,
which is TCXSB with subsumptive tabling for this specific benchmark,
while Alda-generated XSB programs use \co{auto\_table}, which is
variant tabling, %
and these %
are known to cause the observed performance differences~\cite{TekLiu10RuleQuery-PPDP,TekLiu11RuleQueryBeat-SIGMOD}.
\arxiv{%
Alda compiler can be extended to automatically generate optimal
programs using previously studied
methods~\cite{LiuSto09Rules-TOPLAS,Tek+08RulePE-AMAST,TekLiu10RuleQuery-PPDP,TekLiu11RuleQueryBeat-SIGMOD}.}

\item TCpy and TCda, while being drastically easier to write than low-level
  code despite not as easy as rules, are exceedingly inefficient. In
  contrast, TC and TCrev that use rules are drastically faster.
\end{itemize}
Note that both TC and TCrev in Alda, even including the extra times
interfacing with XSB, are faster or even drastically faster than all
systems reported in OpenRuleBench~\cite{Lia+09open} except for XSB and 1 or
2 other systems (for 50K edges, 17 seconds for TC in Alda vs.\ up to 184
seconds and even an error on cyclic data, and 8 seconds for TC in Alda vs.\
up to 120 seconds on acyclic data, where the XSB query times were similar
as in our measurements; note that these are despite OpenRuleBench reporting
only the times for queries, not reading data or writing results).
This is despite all those programs having been manually optimized in the
most advantageous and beneficial ways for each system~\cite{Lia+09open}.

\mysubsec{Integrating with objects, updates, and set queries}
\label{sec-expe-rbac}
To evaluate the performance of using rules with objects and updates, and
of different ways of using rules as well as not using rules, we use the
RBAC benchmarks in Section~\ref{sec-rbac}.

We create 5000 users and 500 roles, %
and randomly generate a user-role assignment \co{UR} of size
5500 %
with a maximum of 10 roles per user, %
and a role hierarchy \co{RH} of size
550 %
and height 5.  %
We run the following set of workloads: iterate and randomly do one of the
following operations in each iteration: add/delete user (50 total each of
add and delete), add/delete role (5 total each), add/delete \co{UR} pair
(55 total each), add/delete \co{RH} pair (5 total each), and query
authorized users ($n$ total), for $n$ up to 500 at intervals of 50.  We
measure the running time of the workload for each $n$.

\oops{}

\pgfplotstableread{
query RBACunion RBACallloc RBACnonloc RBACunionEx RBACalllocEx RBACnonlocEx
 50   33.325   33.448  14.579          13.476     14.575  2.519
100   66.529   66.534  23.865          26.880	  29.022  2.430
150   99.523  100.032  33.591          40.054	  43.608  2.760
200  132.518  133.369  42.837          53.327	  58.314  2.745
250  165.922  166.691  51.969          66.903	  72.703  2.616
300  198.753  199.152  61.801          80.052	  86.409  2.777
350  232.440  233.429  70.418          93.513	 102.006  2.752
400  265.139  266.651  79.941         106.860	 116.418  2.813
450  298.280  298.790  89.045         120.200	 130.043  2.846
500  331.697  333.851  97.923         134.497	 145.940  2.831
}\RBACdata
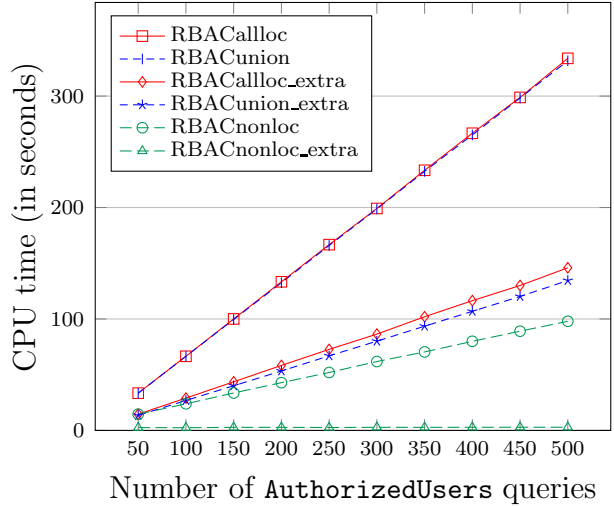
\begin{figure}[t]
  \centering
\begin{tikzpicture}[every mark/.append style={mark size=2pt}]
  \begin{axis}[
    ymin=0, ymax=384,
    xtick=data, xticklabels from table={\RBACdata}{query},
    xlabel={Number of \co{AuthorizedUsers} queries},
    ylabel={CPU time (in seconds)},
    ylabel shift=-1ex,
    tick label style={font=\arxiv{\scriptsize}\oops{}},
    mark options={solid},
    legend style={cells={anchor=west}, legend pos=north west},
    legend style={font=\arxiv{\scriptsize}\oops{}, row sep=-.5ex},
    ymajorgrids=true,
]
\addplot[mark=square,color=red] table[x=query,y=RBACallloc] {\RBACdata};
\addplot[mark=|,color=blue,densely dashed] table[x=query,y=RBACunion] {\RBACdata};
\addplot[mark=diamond,color=red] table[x=query,y=RBACalllocEx] {\RBACdata};
\addplot[mark=star,color=blue,densely dashed] table[x=query,y=RBACunionEx] {\RBACdata};
\addplot[mark=o,color=ForestGreen,dash pattern=on 4.5pt off 2pt] table[x=query,y=RBACnonloc] {\RBACdata};
\addplot[mark=triangle,color=ForestGreen,dash pattern=on 4.5pt off 2pt] table[x=query,y=RBACnonlocEx] {\RBACdata};

\legend{RBACallloc, RBACunion, RBACallloc\_extra, RBACunion\_extra, 
  RBACnonloc, RBACnonloc\_extra}
\end{axis}
\end{tikzpicture}\Vex{-1}
\caption{Running times of RBAC benchmarks, for a workload of updates and
  queries over 5000 users, 500 roles, 5500 user-role assignments, and 550
  role hierarchy pairs.  Running times for RBACpy and RBACda are not shown
  because they are much larger and above all times shown: on data point 50,
  they are \oops{}\arxiv{688.677 and 384.623} seconds, 
  respectively, but they increased linearly 
  as expected,
  to 
  \arxiv{1381.543 seconds on data point 100 and to 1517.510 seconds on data
    point 200}, respectively, and failed to complete by the time limit of
  30 minutes on larger data points.}
\label{fig-RBAC}
\end{figure}

\pgfplotstableread{
query   2pre    Wdata  prepStart  xsbStart xsbRdata xsbWres  2post
 50	 2.935	0.181   2.754	   5.184   0.288     2.708    2.315
100	 5.854	0.386   5.468	  10.417   0.546     5.442    4.595
150	 8.756	0.591   8.165	  15.488   0.814     8.220    6.830
200	11.591	0.750  10.841	  20.725   1.047    11.013    9.123
250	14.672	0.971  13.701	  25.793   1.329    13.736   11.409
300	17.458	1.146  16.312	  31.033   1.604    16.277   13.645
350	20.432	1.387  19.045	  36.237   1.888    19.190   15.911
400	23.338	1.524  21.814	  41.204   2.119    21.865   18.385
450	26.304	1.765  24.539	  46.382   2.389    24.530   20.586
500	29.393	1.915  27.478	  51.987   2.657    27.466   22.982
}\RBACexUnion
\pgfplotstableread{
query   2pre    Wdata  prepStart  xsbStart xsbRdata xsbWres  2post
 50	 2.989	0.281   2.708	   5.224   0.369     2.708    3.285
100	 5.964	0.522   5.442	  10.381   0.729     5.442    6.506
150	 8.986	0.766   8.220	  15.537   1.104     8.220    9.761
200	12.088	1.075  11.013	  20.727   1.454    11.013   13.031
250	15.067	1.331  13.736	  25.878   1.766    13.736   16.256
300	17.825	1.548  16.277	  30.801   2.135    16.277   19.371
350	21.011	1.821  19.190	  36.471   2.526    19.190   22.808
400	24.019	2.154  21.865	  41.387   2.893    21.865   26.256
450	26.814	2.284  24.530	  46.298   3.193    24.530   29.208
500	30.182	2.716  27.466	  51.797   3.590    27.466   32.905
}\RBACexAllloc
\pgfplotstableread{
query   2pre   Wdata  prepStart	 xsbStart xsbRdata xsbWres  2post
 50	0.492  0.033  0.459	 0.963	  0.077	   0.459    0.528
100	0.471  0.034  0.437	 0.926	  0.087	   0.437    0.508
150	0.549  0.042  0.507	 1.084	  0.085	   0.507    0.535
200	0.542  0.036  0.506	 1.058	  0.093	   0.506    0.546
250	0.502  0.033  0.469	 1.027	  0.082	   0.469    0.536
300	0.543  0.037  0.506	 1.065	  0.095	   0.506    0.568
350	0.545  0.04   0.505	 1.079	  0.088	   0.505    0.535
400	0.549  0.039  0.51	 1.077	  0.089	   0.510    0.587
450	0.564  0.048  0.516	 1.101	  0.088	   0.516    0.577
500	0.552  0.035  0.517	 1.094	  0.100	   0.517    0.568
}\RBACexNonloc
\begin{figure}[t]
  \centering
\begin{tikzpicture}[
  every axis/.style={
    ybar stacked, bar width=1ex,
    ymin=0, ymax=189,
    xtick=data, xticklabels from table={\RBACexUnion}{query},
    xlabel={Number of \co{AuthorizedUsers} queries},
    ylabel={CPU time (in seconds)},
    ylabel shift=-1ex,
    tick label style={font=\arxiv{\scriptsize}\oops{}},
    legend style={cells={anchor=west}, legend pos=north west},
    legend style={font=\arxiv{\scriptsize}\oops{}, row sep=-.5ex},
    reverse legend=true,
    ymajorgrids=true, %
    height=40ex, %
    width=0.55\linewidth
  }
  ]
\newcommand{\axisRBACex}[3]{%
\begin{axis}[bar shift=#2]
  \addplot[fill=yellow!#3] table[y=prepStart,meta=query,x expr=\coordindex] {#1};
  \addplot[fill=orange!#3] table[y=Wdata,meta=query,x expr=\coordindex] {#1};

  \addplot[fill=pink!#3] table[y=xsbStart,meta=query,x expr=\coordindex]{#1};
  \addplot[fill=green!#3] table[y=xsbRdata,meta=query,x expr=\coordindex]{#1};
  \addplot[fill=red!#3] table[y=xsbWres,meta=query,x expr=\coordindex] {#1};
  \addplot[fill=blue!#3] table[y=2post,meta=query,x expr=\coordindex] {#1};
  \legend{2pre\_prepStart\!\!,2pre\_rest, xsbStart, xsbRdata\!,xsbWres, 2post}

  \node[draw,fill=white,anchor=north east] at (rel axis cs: 0.86,0.97){\shortstack[l]{
      {\arxiv{\scriptsize}\oops{} \Hex{-1}{\bf Bar triples:}}\\
      {\arxiv{\scriptsize}\oops{} \Hex{-1}left: RBACallloc\_extra}\\
      {\arxiv{\scriptsize}\oops{} \Hex{-1}middle: RBACunion\_extra}\\
      {\arxiv{\scriptsize}\oops{} \Hex{-1}right: RBACnonloc\_extra}
    }};
\end{axis}
}
\axisRBACex{\RBACexNonloc}{1ex}{60}%
\axisRBACex{\RBACexUnion}{0ex}{60}%
\axisRBACex{\RBACexAllloc}{-1ex}{60}
\end{tikzpicture}\Vex{-1}
\caption{Breakdown of RBACallloc\_extra, RBACunion\_extra, and
  RBACnonloc\_extra.}
\label{fig-RBAC-break}
\end{figure}
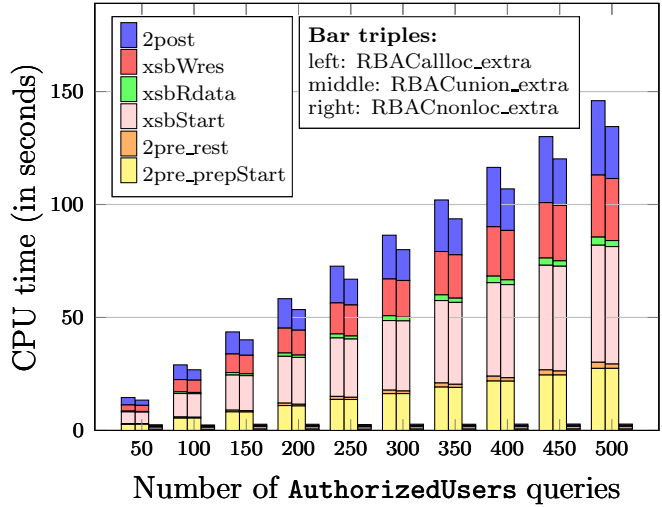

Figure~\ref{fig-RBAC} shows the running times of the RBAC benchmarks, all
scaling linearly in the number of \co{AuthorizedUsers} queries, as
expected.
Labels with suffix \_extra indicate the part of the running time of the
corresponding program for extra work interacing with XSB: 2pre, 2post,
xsbRdata, and xsbWres as in Figure~\ref{fig-TC-break} plus here the times
for starting XSB processes.

Figure~\ref{fig-RBAC-break} shows the breakdown of the times for the extra
work.  It also highlights the part of 2pre on preparing the queries and
commands to start XSB
(2pre\_prepStart), with the rest of 2pre (2pre\_rest) on writing data to
files for XSB to read.  xsbStart is the time for starting the XSB process.

We observe that:
\begin{itemize}

\item The extra times interfacing with XSB are again obvious, %
  but here dominated mostly by preparing queries and commands and starting
  XSB, as shown in Figure~\ref{fig-RBAC-break}, unlike for TC benchmarks,
  because the data and results are much smaller but all the work associated
  with invoking XSB through command line is repeated \m{n} (50 to 500)
  times.  
  This overhead from going through files will also be removed when using an
  in-memory interface between XSB and Python without starting XSB
  repeatedly.
  
\item RBACallloc and RBACunion are very close, as shown in
  Figure~\ref{fig-RBAC}, with a slightly higher interfacing overhead by
  RBACallloc as expected for the extra data and results passed due to
  \co{ROLES}, but compensated by a slightly faster queries in XSB than set
  operations in Python.
  RBACnonloc is more than 3 times as fast as RBACallloc and RBACunion
  and is the fastest, %
  because the inference for computing \co{transRH} is done at updates not
  queries, and there is a smaller, fixed number of updates.
  Its performance can be optimized even more with incremental computation,
  as for either set queries,
  e.g.,~\cite{Liu+06ImplCRBAC-PEPM,Gor+12Compose-PEPM,Liu+16IncOQ-PPDP}, or
  logic rules, e.g.,~\cite{SahaRam03}.

\item RBACpy and RBACda are again exceedingly inefficient, as expected.
  In contrast, the three programs that use rules are all significantly
  faster.
\end{itemize}

\mysubsec{Integrating with aggregate queries and recursive functions}
\label{sec-expe-pa}

To evaluate the performance of integrated use of rules with aggregate
queries and recursive functions, we use two benchmarks for class hierarchy
analysis: PA and PAopt, and the corresponding programs in XSB, as described
in Section~\ref{sec-pa}.

We found the XSB programs corresponding to PA and PAopt, which we call
PAXSB and PAoptXSB, respectively, to be highly inefficient, being slower
and even drastically slower than Alda programs.
We tried many manual optimizations by manipulating the rules and adding
directives, including with help from an XSB expert, and selected the best
version, which we call PAXSBopt, that uses additional directives for
targeted tabling that also subsumes some indexing.

The programs analyzed %
include 9 widely-used open-source Python packages, %
all available on GitHub (\myurl{https://github.com/}):
NumPy (\ver{v1.21.5}\ghurl{https://github.com/}{numpy/numpy}) and %
SciPy (\ver{v1.7.3}\ghurl{https://github.com/}{scipy/scipy}), for scientific computation;
MatPlotLib (matplot) (\ver{v3.5.1}\ghurl{https://github.com/}{matplotlib/matplotlib}), for visualization;
Pandas (\ver{v1.3.5}\ghurl{https://github.com/}{pandas-dev/pandas}), for data analysis; 
SymPy (\ver{sympy-1.9}\ghurl{https://github.com/}{sympy/sympy}), for symbolic computation; 
Django (\ver{4.0}\ghurl{https://github.com/}{django/django}), for web development; 
Scikit-learn (sklearn) (\ver{1.0.1}\ghurl{https://github.com/}{scikit-learn/scikit-learn}) 
and
PyTorch (\ver{v1.10.1}\ghurl{https://github.com/}{pytorch/pytorch}), for machine learning;
and 
Blender (\ver{v3.0.0}\ghurl{https://github.com/}{blender/blender}), for 3D graphics.
Each of these Python packages %
contains many files and directories.
We first parse each file and translate the resulting abstract
syntax tree (AST) along with file and directory information into Datalog
facts.
We then run the benchmarks.

\begin{table*}[t]
  \footnotesize
  \centering
  \begin{tabular}{@{\,}l@{\,} @{\,}l@{\,} ||@{\,}r@{\,}|@{\,}r@{\,}|@{\,}r@{\,}|@{\,}r@{\,}|@{\,}r@{\,}|@{\,}r@{\,}|@{\,}r@{\,}|@{\,}r@{\,}|@{\,}r@{\,}}
Measure & Item/Name     & numpy   & django  & sklearn & blender & pandas  & matplot & scipy   & pytorch   & sympy \\
    \hline\hline
Data	& Total	        & 640,715 & 815,551 & 862,031 & 909,600 & 942,315 & 1,064,859 & 1,092,466 & 5,142,905 & 5,115,105 \\
size	& \co{ClassDef}	& 587	  & 1,835   & 535     & 2,146	& 849	  & 994	    & 898     & 6,467	  & 1,830 \\
	& \co{Name}	& 96,076  & 119,077 & 137,066 & 107,638	& 153,664 & 152,357 & 178,754 & 797,072	  & 1,063,842 \\
 	& \co{Member}	& 155,207 & 199,416 & 210,410 & 242,531	& 227,766 & 268,736 & 260,848 & 1,270,917 & 1,112,296 \\
 	& Total used    & 251,870 & 320,328 & 348,011 & 352,315 & 382,279 & 422,087 & 440,500 & 2,074,456 & 2,177,968 \\\hline
Ratio   & Used/Total    & 39.3\%  & 39.3\%  & 40.4\%  & 38.7\%  & 40.6\%  & 39.6\%  & 40.3\%  & 40.3\%    & 42.6\% \\
    \hline\hline

Result  & \co{\footnotesize \#defined}       & 519 &  1610 &   533 &  2118 &   804 &   935 &   882 &  4323 &  1786 \\
size    & \co{\footnotesize \#extending}     & 419 &  1457 &   710 &  2951 &   407 &   610 &   719 &  2207 &  1816 \\
        & \co{\footnotesize \#roots}         &  79 &   225 &    51 &   133 &    88 &   104 &    60 &   137 &    92 \\
        & \co{\footnotesize max\_height}     &   8 &     7 &     5 &     4 &     7 &     5 &     3 &     5 &    12 \\
        & \co{\footnotesize \#roots\_max\_h} &   1 &     2 &     2 &     1 &     2 &     1 &     4 &     1 &     1 \\
        & \co{\footnotesize \#desc}          & 427 &  2329 &   822 &  4376 &   436 &   605 &   721 &  2174 &  2413 \\
        & \co{\footnotesize max\_desc}       &  84 &   309 &   256 & 1,638 &    65 &    47 &   353 & 1,045 & 1,078 \\
        & \co{\footnotesize \#roots\_max\_d} &   1 &     1 &     1 &     1 &     1 &     1 &     1 &     1 &     1 \\
    \hline\hline

Running & PA       &  2.542 &   3.573 &  3.263 &   5.134 &  3.342 &  3.733 &  3.646 &  14.652 &  15.243 \\
time    & PAopt    &  2.631 &   3.520 &  3.235 &   4.661 &  3.341 &  3.676 &  3.633 &  14.706 &  15.132 \\
(in     & PAXSB    &  6.297 & 112.091 & 10.795 & 243.765 &  6.378 & 14.400 & 22.221 & 969.228 &  65.382 \\
seconds)& PAoptXSB & 13.170 & 343.428 & 17.066 & 326.629 & 18.863 & 40.871 & 29.675 &1773.374 & 181.961 \\
        & PAXSBopt &  0.804 &   1.499 &  1.071 &   2.149 &  1.118 &  1.317 &  1.300 &   5.158 &   5.051 \\
    \hline
Ratio	& PAopt    &103.5\% &  98.5\% & 99.1\% &  90.8\% &100.0\% &  98.5\% & 99.6\% &  100.4\% &  99.3\% \\
over	& PAXSB    &247.7\% &3137.2\% &330.8\% &4748.0\% &190.8\% & 385.7\% &609.5\% & 6615.0\% & 428.9\% \\ 
PA	& PAoptXSB &518.1\% &9611.7\% &523.0\% &6362.1\% &564.4\% &1094.9\% &813.9\% &12103.3\% &1193.7\% \\
	& PAXSBopt & 31.6\% &  42.0\% & 32.8\% &  41.9\% & 33.5\% &  35.3\% & 35.6\% &   35.2\% &  33.1\% \\
    \hline\hline
  \end{tabular}\Vex{1}
  \caption{Data sizes, analysis results, and running times for program
    analysis benchmarks.  Total is the total number of facts about each
    package.  Total used is the sum of numbers of \co{ClassDef}, \co{Name},
    and \co{Member} facts.}
  \label{tab-PA-expe}
\end{table*}
Table~\ref{tab-PA-expe} shows data sizes,
analysis results, and running times of the analysis.
The columns are sorted by the total number of facts used (i.e., all
\co{ClassDef}, \co{Name}, and \co{Member} facts), which mostly coincides
with the total number of facts except for the largest two, \co{pytorch} and
\co{sympy}.
We can see that, even for the small rule set \co{class\_extends\_rs}, the
total number of facts used is already 38.7-42.6\% of the total number of
facts, because \co{Member} and \co{Name} are two of the largest.

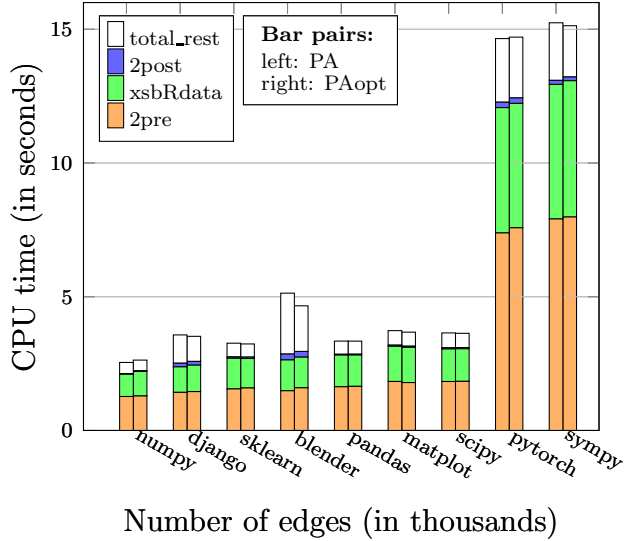
\begin{figure}[t]
  \centering
\begin{tikzpicture}[
  every axis/.style={
    ybar stacked, bar width=1ex,
    ymin=0, ymax=16,
    xtick=data, xticklabels from table={\PAex}{lib},
    xlabel={Number of edges (in thousands)},
    ylabel={CPU time (in seconds)},
    ylabel shift=-1ex,
    tick label style={font=\arxiv{\scriptsize}\oops{}},
    x tick label style={at={(axis description cs:2.01,+0.0)},rotate=-30,anchor=west},
    legend style={cells={anchor=west}, legend pos=north west},
    legend style={font=\arxiv{\scriptsize}\oops{}, row sep=-.5ex},
    reverse legend=true,
    ymajorgrids=true, %
  }
  ]
\newcommand{\axisPAex}[3]{%
\begin{axis}[bar shift=#2]
  \addplot[fill=orange!#3] table[y=2pre,meta=lib,x expr=\coordindex] {#1};
  \addplot[fill=green!#3] table[y=xsbRdata,meta=lib,x expr=\coordindex]{#1};
  \addplot[fill=blue!#3] table[y=2post,meta=lib,x expr=\coordindex] {#1};
  \addplot[fill=white!#3] table[y=rest,meta=lib,x expr=\coordindex] {#1};
  \legend{2pre, xsbRdata\!, 2post, total\_rest}
  \node[draw,fill=white,anchor=north east] at (rel axis cs: 0.61,0.97){\shortstack[l]{
      {\arxiv{\scriptsize}\oops{} {\bf Bar pairs:}}\\
      {\arxiv{\scriptsize}\oops{} left: PA}\\
      {\arxiv{\scriptsize}\oops{} right: PAopt}
    }};
\end{axis}
}
\axisPAex{\PAoptex}{1ex}{60}%
\axisPAex{\PAex}{0ex}{60}
\end{tikzpicture}\Vex{-1}
\caption{Running times of PA and PAopt.}
\label{fig-PA-break}
\end{figure}

Figure~\ref{fig-PA-break} shows the breakdown of the time interfaceing with
XSB (2pre, 2post, and xsbRdata, as in Figure~\ref{fig-TC-break} except that
xsbWres is even smaller than 2post and is not shown) plus the remaining
time in the total time for PA and PAopt (total\_rest).
\pgfplotstableread{
lib  2pre  2post xsbRdata     total   rest
numpy    1.270 0.025 0.833     2.542  0.415
django   1.425 0.137 0.953     3.573  1.058
sklearn  1.553 0.051 1.147     3.263  0.512
blender  1.485 0.220 1.152     5.134  2.277
pandas   1.634 0.030 1.185     3.342  0.493
matplot  1.832 0.043 1.315     3.733  0.543
scipy    1.831 0.046 1.216     3.646  0.554
pytorch  7.386 0.204 4.684    14.652  2.378
sympy    7.908 0.152 5.030    15.243  2.153
}\PAex
\pgfplotstableread{
lib  2pre  2post xsbRdata     total   rest
numpy    1.289 0.027 0.917     2.631  0.398
django   1.454 0.14  0.988     3.52   0.938
sklearn  1.589 0.053 1.102     3.235  0.491
blender  1.595 0.213 1.146     4.661  1.707
pandas   1.650 0.032 1.170     3.341  0.489
matplot  1.791 0.042 1.318     3.676  0.525
scipy    1.837 0.045 1.218     3.633  0.533
pytorch  7.578 0.205 4.651    14.706  2.272
sympy    7.981 0.143 5.094    15.132  1.914
}\PAoptex
We can see that:
\begin{itemize}
\item The times interfacing with XSB is again obvious, here vastly
  dominated by the time to pass AST facts to XSB as shown in
  Figure~\ref{fig-PA-break}, because of the large data sizes vs.\ the small
  result sizes shown in Table~\ref{tab-PA-expe}.  This contrasts the times
  dominated by passing results in Figure~\ref{fig-TC-break} and by repeated
  starting of XSB in Figure~\ref{fig-RBAC-break}; XSB is invoked only twice
  here in each run, once for each rule set in the benchmark.  Again, this
  overhead will be removed by in-memory mapping between XSB and Python data
  structures.

\item The running times of PA and PAopt are similar and mostly increase as
  the data sizes increase, as shown in Table~\ref{tab-PA-expe} and
  Figure~\ref{fig-PA-break}.
  PAopt is in most cases (all but \co{numpy} and \co{pytorch}) very
  slightly faster than PA, because querying using rules takes only a small
  part of the total time (0.5--11.6\% of PA, and 0.3--2.5\% of PAopt), with
  the rest on interfacing with XSB and on other queries using functions and
  aggregations.
  Querying using rules in PAopt is actually 1.4--11.8 times as fast as that
  in PA.

\item PAXSB and PAoptXSB have vastly varying running times, as shown in
  Table~\ref{tab-PA-expe}, unlike PA and PAopt, and are much slower than PA
  and PAopt, taking 1.9--121.1 times as long as PA and longer than PAopt.
  This is after we already manually added tabling for \co{height} and
  \co{num\_desc} to match PA and PAopt, after finding that \co{auto\_table}
  only tabled predicate \co{desc}.
  We can see that PAXSB and PAoptXSB are mostly slower for larger result
  sizes, as opposed to input sizes, though all result sizes are orders of
  magnitude smaller than input sizes.

  PAXSBopt, with manual optimizations after trying various combinations of
  tabling, indexing, and rewriting for the remaining predicates, is
  58.0-78.4\%
  faster that PA.  Again, previously studied
  methods~\cite{TekLiu10RuleQuery-PPDP,TekLiu11RuleQueryBeat-SIGMOD} can be
  added to the Alda compiler to automatically generate optimal tabling and
  indexing directives as needed; manually applying these sophisticated
  methods is too tedious.

\end{itemize}

\mysubsec{Scaling with data and rules}
\label{sec-expe-scale}

We examine how the performance scales for large sizes of data and rules
using two benchmarks in OpenRuleBench:
DBLP, the last under large join tests, with the largest real-world data set
among all benchmarks in OpenRuleBench; and Wine, the last under Datalog
recursion, with the largest rule set among all benchmarks in OpenRuleBench.
Again, we changed \co{load\_dyn} used in OpenRuleBench to \co{load\_dync}
for faster reading of facts in XSB's canonical form.

The DBLP benchmark does a 5-way join with projections, on DBLP data containing 2.4+ million facts.
The Wine benchmark in OpenRuleBench has 961 rules and 654 facts; it was
originally too slow in XSB but optimized using subsumptive
transformations~\cite{TekLiu11RuleQueryBeat-SIGMOD}, resulting in 967
rules.  The Wine benchmark in Alda is translated from the optimized rules.

Table~\ref{tab-dblp-wine} shows the running times for DBLP and Wine
benchmarks, for both the Alda programs and the XSB programs.
\_extra under Alda is the part of the total time on 2pre, 2post, xsbRdata,
and xsbWres.  OrigTotal under XSB is the Total time for the original
program from OpenRuleBench, which uses \co{load\_dyn} instead of
\co{load\_dync}.  We note that:
\begin{table}
  \small
  \centering
\begin{tabular}{@{\m}l@{\,}||r@{~}|@{~}r||@{~}r@{~}|@{~}r@{~}|@{~}r@{~}|@{~}r@{~}|@{~}r@{~}||@{~}c@{~}||c@{~}|@{\,}r@{\,}}
       & \multicolumn{8}{c||}{Alda} & \multicolumn{2}{c}{XSB}\\
  \cline{2-11} 
  Name & RawR & PickleW %
                      &2pre &xsbRdata &xsbWres &2post &\_extra~&Total &Total 
& OrigTotal \\\hline
  DBLP &12.187 &3.131 &15.722 &11.197 &0.054 &0.020 &26.891 & 30.573 & 9,492
& 63.494\\\hline
  Wine & 0.008 &0.000 & 0.037 & 0.219 &0.000 &0.001 & 0.213 & 30.960 & 3.754
&  3.826\\\hline
\oops{}
\end{tabular}
\caption{Running times (in seconds) of DBLP and Wine benchmarks. 
  \oops{}\oops{}}
\label{tab-dblp-wine}
\end{table}

\begin{itemize}
\item
For the DBLP benchmark, XSB is more than 3 times as fast as Alda.  The
large data size causes 2pre and xsbRdata to dominate the interfacing
overhead, as for PA and PAopt benchmarks.
\oops{}%
Again, this overhead of going through files will be removed with an
in-memory interface.

Alda's reading of raw data (12.187 seconds) is higher than XSB's for DBLP
due to the use of Python regular expressions to parse extra string formats
while XSB benefits from drastically reduced checks reading their canonical
data form. This can be fixed with a specialized reading function in C
similar to the one used by XSB.

Note that the original XSB benchmark from OpenRuleBench (63.494 seconds),
without our optimization to use faster data loading, is much slower than
even Alda (30.573 seconds) that includes the extra overhead.  
Because OpenRuleBench reports only query time, which is small (about 1.3
second using XSB from Alda) compared with reading large data for DBLP, our
total time that includes reading data and writing results is larger than
all times reported in OpenRuleBench~\cite{Lia+09open}, but XSB was the
second fastest there, and one system produced an error.

\item 
For the Wine benchmark,\oops{}
XSB is more than 8 times as fast as Alda, due to the use of \co{auto\_table}
in Alda generated code.  Manually added subsumptive tabling in the XSB
benchmark from OpenRuleBench reduces the XSB query time from 27.722 seconds
to 3.747 seconds.  Again the Alda compiler can be extended with automatic
optimizations~\cite{LiuSto09Rules-TOPLAS,TekLiu10RuleQuery-PPDP,TekLiu11RuleQueryBeat-SIGMOD}.

Note that Alda is still faster than half of the systems tested in
OpenRuleBench, up to 140 seconds and even errors in three systems, where
XSB was the fastest at 4.47 seconds~\cite{Lia+09open}.
\oops{}
\end{itemize}

\section{Related work and conclusion}
\label{sec-related}

There has been extensive effort in the design and implementation of
languages to support programming with logic rules together with other
programming paradigms, by extending logic languages, extending languages in
other paradigms, or developing multi-paradigm or other standalone
languages.

A large variety of logic rule languages have been extended to support
sets, functions, updates, and/or objects, etc.  For example, see Maier et
al.~\cite{maier18hist-wbook} for Datalog and variants extended with sets,
functions, objects, updates, higher-order extensions, and more.
In particular, many Prolog variants support sets, functions, updates, 
objects, constraints, etc.  For example, Prolog supports \co{assert} for
updates, as well as cut and negation as failure that are imperative instead
of declarative;
Flora~\cite{YanKif00flora,flora20} builds on XSB and supports objects
(F-logic), higher-order programming (HiLog), and updates (Transaction
Logic); and Picat~\cite{Zhou16picat} builds on B-Prolog and
supports updates, comprehensions, etc.
Lambda Prolog~\cite{miller2012programming} extends Prolog with 
simply typed lambda terms and higher-order programming.
Functional logic languages, such as Mercury~\cite{somogyi1995mercury} 
and Curry~\cite{hanus2013functional}, 
combine functional programming and logic programming.
Some logic programming systems are driven by scripting externally, e.g.,
using Lua for IDP~\cite{bru14predicate}, and shell scripts for
LogicBlox~\cite{aref2015design}.
Flix~\cite{madsen2016datalog,madsen2020fixpoints} extends 
Datalog with lattices and monotone functions, and functional programming.
These languages are intrinsically driven by logic rules or functional
programming, and do not support commonly-used updates, objects, and sets
in a simple and direct way as in a general powerful language like Alda, or
do not support some or all of them at all.

Many languages in other programming paradigms, especially imperative
languages and object-oriented languages, have been extended to support
rules.
This is notably through explicit interfaces with particular logic
languages, for examples, a Java interface for XSB through
InterProlog~\cite{calejo04inteprolog,xsb21}, C++ and Python interfaces for
answer-set programming systems dlvhex~\cite{Redl16} and
Potassco~\cite{ban17clingcon}, and a Python interface for
IDP~\cite{ven17idp-py}.
While imperative and object-oriented languages support easy and direct
updates and object encapsulation, interfacing
with logic languages through 
explicit interfaces requires programmers to write tedious, low-level
wrapper code for going to the rule language and coming back.  They are in
the same spirit as interfaces such as JDBC~\cite{reese2000database} for
using database systems from languages such as Java.

Multi-paradigm languages and other standalone languages 
have also been developed.  For example, the Mozart %
system for the Oz multi-paradigm programming
language~\cite{RoyHar04} supports logic, functional, and
constraint as well as imperative and concurrent programming. However, it is
similar to logic languages extended with other features, because it supports
logic variables, but not state variables to be assigned to as in commonly-used imperative
languages.
Examples of other languages involving logic and constraints with updates
include TLA+~\cite{lam94tla}, a logic language for
specifying actions;
CLAIRE~\cite{caseau2002claire}, an object-oriented language that supports
functions, sets, and rules whose conclusions are actions;
LINQ~\cite{meijer2006linq,LINQ}, an extension of C\# for SQL-like queries;
IceDust~\cite{harkes16IceDust}, a Java-based language for querying data
with path-based navigation and incremental computation;
extended LogiQL in SolverBlox~\cite{aref18solverblox-wbook}, for
mathematical and logic programming on top of Datalog with updates and
constraints; and other logic-based query languages,
e.g.,~Datomic~\cite{anderson2016datomic} and SOUL~\cite{de2011soul}.
These are either logic languages lacking general imperative and
objected-oriented programming constructs, or imperative and object-oriented
languages lacking the power of logic rules.

In conclusion, Alda allows the use of logic rules with all of sets,
functions, updates, and objects in a seamlessly integrated fashion.
As a direction for future work, many optimizations can be used to improve
the efficiency of implementations.
This includes optimizing the logic rule engines used, improving interfaces
and interactions with them, and using different and specialized rule
engines such as Souffle~\cite{jordan2016souffle} to obtain the best
possible performance.

\mypar{\large Acknowledgments}
We would like to thank David S.\ Warren for a 28-line XSB program as an initial implementation of the external interface to XSB.
We thank students in undergraduate and graduate courses for using earlier versions of Alda, called DA-rules, and Thang Bui in particular for additional applications using Alda in program analysis and optimization and for help supervising some other students.

\def\usebib{
\def\bibdir{../../../bib}         %
{
\renewcommand{\baselinestretch}{-.1}\small%
\bibliography{\bibdir/strings,\bibdir/liu,\bibdir/IC,\bibdir/PT,\bibdir/PA,\bibdir/Lang,\bibdir/Algo,\bibdir/DB,\bibdir/AI,\bibdir/Sec,\bibdir/Sys,\bibdir/SE,\bibdir/Vis,\bibdir/misc,\bibdir/crossref} %
\arxiv{
\bibliographystyle{alpha}
}
}
}
\newcommand{\etalchar}[1]{$^{#1}$}

\appendix
\oops{}

\newcommand{\twocolonly}[1]{#1}






\section{Formal Semantics}
\label{app-formal}

\renewcommand{\topfraction}{0.99}	
\renewcommand{\bottomfraction}{0.99}	
\renewcommand{\textfraction}{0.01}	
\renewcommand{\dbltopfraction}{0.99}	
\renewcommand{\dblfloatpagefraction}{0.9}
\renewcommand{\floatpagefraction}{0.9}
\setcounter{topnumber}{2}
\setcounter{bottomnumber}{2}
\setcounter{totalnumber}{4}     
\setcounter{dbltopnumber}{2}

\newenvironment{ctabbing}
          {\begin{center}\begin{minipage}{\textwidth}\begin{tabbing}}
          {\end{tabbing}\end{minipage}\end{center}}

\newcommand{\firstalt}{~~}
\newcommand{\alt}{~~}

\newcommand{\negspc}{}

\newcommand{\pfn}{\rightharpoonup}
\newcommand{\bijection}{\stackrel{1-1}{\rightarrow}}
\newcommand{\union}{\cup}
\newcommand{\intersect}{\cap}
\newcommand{\UNION}{\bigcup}
\newcommand{\LAND}{\bigwedge}
\newcommand{\LOR}{\bigvee}
\newcommand{\Set}[1]{{\rm Set}(#1)}
\newcommand{\ra}{\rightarrow}
\newcommand{\myiff}{\Leftrightarrow}

\newcommand{\Bool}{{\it Bool}}
\newcommand{\Int}{{\it Int}}
\newcommand{\Address}{\mathify{\it Address}}
\newcommand{\ProcessAddress}{{\it ProcessAddress}}
\newcommand{\NonProcessAddress}{{\it NonProcessAddress}}
\newcommand{\Val}{\mathify{\it Val}}
\newcommand{\Object}{{\it Object}}
\newcommand{\MsgQueue}{{\it MsgQueue}}
\newcommand{\ChannelStates}{{\it ChannelStates}}
\newcommand{\LocalHeap}{{\it LocalHeap}}
\newcommand{\Heap}{{\it Heap}}
\newcommand{\HeapType}{{\it HeapType}}
\newcommand{\State}{{\it State}}
\newcommand{\Program}{\mathify{\it Program}}
\newcommand{\Configuration}{\mathify{\it Configuration}}
\newcommand{\ProcessClass}{\mathify{\it ProcessClass}}
\newcommand{\Method}{\mathify{\it Method}}
\newcommand{\ReceiveDef}{\mathify{\it ReceiveDef}}
\newcommand{\ReceivePattern}{\mathify{\it ReceivePattern}}
\newcommand{\Pattern}{\mathify{\it Pattern}}
\newcommand{\InstanceVariable}{\mathify{\it InstanceVariable}}
\newcommand{\MethodName}{\mathify{\it MethodName}}
\newcommand{\Parameter}{\mathify{\it Parameter}}
\newcommand{\Expression}{\mathify{\it Expression}}
\newcommand{\Iterator}{\mathify{\it Iterator}}
\newcommand{\AnotherAwaitClause}{\mathify{\it AnotherAwaitClause}}
\newcommand{\Literal}{\mathify{\it Literal}}
\newcommand{\BooleanLiteral}{\mathify{\it BooleanLiteral}}
\newcommand{\IntegerLiteral}{\mathify{\it IntegerLiteral}}
\newcommand{\UnaryOp}{\mathify{\it UnaryOp}}
\newcommand{\BinaryOp}{\mathify{\it BinaryOp}}
\newcommand{\TuplePattern}{\mathify{\it TuplePattern}}
\newcommand{\PatternElement}{\mathify{\it PatternElement}}
\newcommand{\ChannelOrder}{\mathify{\it ChannelOrder}}
\newcommand{\ChannelReliability}{\mathify{\it ChannelReliability}}
\newcommand{\EC}{\mathify{\it C}}

\newcommand{\ClassName}{\mathify{\it ClassName}}
\newcommand{\Label}{\mathify{\it Label}}
\newcommand{\Field}{\mathify{\it Field}}
\newcommand{\Statement}{\mathify{\it Statement}}
\newcommand{\Tuple}{\mathify{\it Tuple}}

\newcommand{\commentMark}{/\,/}

\newcommand{\commentS}[1]{\mbox{\commentMark\ #1}}

\newcommand{\tuple}[1]{(#1)}
\newcommand{\ltuple}{(}
\newcommand{\rtuple}{)}
\newcommand{\seq}[1]{\langle#1\rangle}
\newcommand{\set}[1]{\{#1\}}

\newcommand{\new}{{\it new}}
\newcommand{\emptySet}{\set{}}
\newcommand{\emptyfn}{f_0}
\newcommand{\emptyseq}{\seq{}}
\newcommand{\IF}{\mbox{if }}
\newcommand{\dom}{{\it dom}}
\newcommand{\range}{{\it range}}
\newcommand{\rest}{{\it rest}}
\newcommand{\length}{{\it length}}
\newcommand{\first}{{\it first}}
\newcommand{\iscopy}{{\it isCopy}}
\newcommand{\matchRcvDef}{{\it matchRcvDef}}
\newcommand{\rcvAtLabel}{{\it receiveAtLabel}}
\newcommand{\rcvMsg}{{\it rcvMsg}}
\newcommand{\bottom}{\mathord{\perp}}
\newcommand{\extends}{{\it extends}}
\newcommand{\methodDef}{{\it methodDef}}
\newcommand{\addrs}{{\it addrs}}
\newcommand{\subst}{{\it subst}}
\newcommand{\Init}{{\it Init}}
\newcommand{\tpli}{\hspace*{0.4em}}
\newcommand{\sci}{\hspace*{0.75em}}
\newcommand{\spce}{\hspace*{1em}}
\newcommand{\hole}{[\hspace*{0.1em}]}

\newcommand{\Class}{\mathify{\it Class}}
\newcommand{\Ruleset}{\mathify{\it Ruleset}}
\newcommand{\RulesetName}{\mathify{\it RulesetName}}
\newcommand{\Rule}{\mathify{\it Rule}}
\newcommand{\Assertion}{\mathify{\it Assertion}}
\newcommand{\Predicate}{\mathify{\it Predicate}}
\newcommand{\PredicateArg}{\mathify{\it PredicateArg}}
\newcommand{\KeywordArg}{\mathify{\it KeywordArg}}
\newcommand{\GlobalVariable}{\mathify{\it GlobalVariable}}
\newcommand{\LogicVariable}{\mathify{\it LogicVariable}}
\newcommand{\Variable}{\mathify{\it Variable}}
\newcommand{\Query}{\mathify{\it Query}}

\newcommand{\agv}{a_{\it gv}}
\newcommand{\pat}{{\it pat}}
\newcommand{\none}{{\tt None}}
\newcommand{\kwargs}{{\it kwargs}}
\newcommand{\slice}{{\it slice}}
\newcommand{\knownbps}{{\it knownBasePreds}}
\newcommand{\evalrules}{{\it evalRules}}
\newcommand{\rulesetsg}{{\it glblRulesets}}
\newcommand{\rulesets}{{\it rulesets}}
\newcommand{\basePredParams}{{\it basePredParams}}
\newcommand{\basePredVars}{{\it basePredVars}}
\newcommand{\knownVars}{{\it knownVars}}
\newcommand{\derivedPredVars}{{\it derivedPredVars}}
\newcommand{\derivedPredParams}{{\it derivedPredParams}}
\newcommand{\derivedPredVarsg}{{\it glblDerivedPredVars}}
\newcommand{\rules}{{\it rules}}
\newcommand{\predicate}{{\it predicate}}
\newcommand{\pattern}{{\it pattern}}
\newcommand{\stk}{{\it stk}}  
\newcommand{\args}{{\it args}}
\newcommand{\freshaddrs}{{\it freshAddrs}}
\newcommand{\newfn}{{\it newFn}}  
\newcommand{\maintain}{{\it maintain}}
\newcommand{\maintsub}{{\it maintSub}}  
\newcommand{\infsub}{{\it infSub}}
\newcommand{\updatevar}{{\it updateVar}}
\newcommand{\legalassign}{{\it legalAssign}}
\newcommand{\result}{{\it result}}
\newcommand{\thetaht}{\theta_{T}}


We give an abstract syntax and formal semantics for a core language for Alda.  It builds on the standard least fixed-point semantics for Datalog~\cite{fitting2002fixpoint}
and the formal operational semantics for DistAlgo~\cite{Liu+17DistPL-TOPLAS}.
We removed the constructs specific to distributed algorithms and added the constructs described in this paper.  The removed DistAlgo constructs can easily be restored to obtain a semantics for the full language; we removed them simply to avoid repeating them.  The operational semantics is a reduction semantics with evaluation contexts \cite{wright94syntactic,serbanuta07rewriting}.

In addition to introducing an abstract syntax for rule sets and calls to \co{\INFER}, and a transition rule for calls to  \co{\INFER}, we extended the state with a stack that keeps track of rule sets whose results need to be maintained, extended several existing transition rules to perform automatic maintenance of the results of rule sets, and modified the semantics of existential quantifiers to bind the quantified variables to a witness when one exists.

\subsection{Abstract syntax}
\label{sec:syntax}

The abstract syntax is defined in Figures \ref{fig:syntax1} and \ref{fig:syntax2}.
Tuples are treated as immutable values, not as mutable objects. Sets and sequences are treated as objects, because they are mutable.  These are predefined classes that should not be defined explicitly.  Methods of {\tt set} include {\tt add}, {\tt any} (which returns an element of the set, if the set is non-empty, otherwise it returns \none), {\tt contains}, {\tt del},  and {\tt size}.  Methods of {\tt sequence} include {\tt add} (which adds an element at the end of the sequence), {\tt contains}, and {\tt length}.  For brevity, among the standard arithmetic operations, we include only one representative operation in the abstract syntax and semantics; others are handled similarly.  All expressions are side-effect free.  Object creation, comprehension, and {\tt infer} are statements, not expressions, because they have side-effects; comprehension has the side-effect of creating a new {\tt set}. Semantically, the {\tt for} loop copies the contents of a (mutable) sequence or set into an (immutable) tuple before iterating over it, to ensure that changes to the sequence or set by the loop body do not affect the iteration.  {\tt whileSome} and {\tt ifSome} are similar to {\tt while} and {\tt if}, except that they always have an existential quantification as their condition, and they bind the variables in the pattern in the quantification to a witness, if one exists.  The literal $\none$ is used to represent ``undefined''.  We use some syntactic sugar in sample code, e.g., we use infix notation for some binary operators, such as {\tt and} and {\tt is}.



\begin{figure*}[htb]
\hspace*{1em}
\begin{ctabbing}
\Program\ ::= \Ruleset* \Class*\ \Statement\\
\Ruleset\ ::= {\tt rules} \RulesetName\ \Rule+\\
\Rule\ ::= \Assertion\ {\tt if} \Assertion*\\
\Assertion\ ::= \Predicate(\PredicateArg*)\\
\Predicate\ ::= \= \GlobalVariable\\
\> {\tt self}.\Field\\
\> \Parameter\\
\PredicateArg\ ::= \= \LogicVariable\\
\> \Literal \\
\Class\ ::= {\tt class} \ClassName\ [{\tt extends} \ClassName]: \Ruleset* \Method*\\
\Method\ ::= \=
 {\tt def} \MethodName{\tt (}\Parameter*{\tt )} \Statement\\
\> {\tt defun} \MethodName{\tt (}\Parameter*{\tt )} \Expression\\
\\
\Statement\ ::= \= \Variable\ {\tt :=} \Expression\\
\> \Variable\ {\tt :=} {\tt new} \ClassName\\
\> \Variable\ {\tt := \{} \Pattern\ {\tt :} \Iterator* {\tt |} \Expression\ {\tt \}}\\
\> \Statement\ {\tt ;} \Statement\\
\> {\tt if} \Expression{\tt :} \Statement\ {\tt else:} \Statement\\
\> {\tt for} \Iterator{\tt :} \Statement\\
\> {\tt while} \Expression{\tt :} \Statement\\
\> {\tt ifSome} \Iterator {\tt |} \Expression\ {\tt :} \Statement\\
\> {\tt whileSome} \Iterator\ {\tt |} \Expression\ {\tt :} \Statement\\
\> \Expression{\tt .}\MethodName{\tt (}\Expression*{\tt )}\\
\> \Variable*\ := [\Expression.]{\tt infer}(\Query*, \KeywordArg*, {\tt rules}=\RulesetName)\\
\> {\tt skip}\\
\Expression\ ::= \=
\Literal\\
\> \Parameter\\
\> \Variable\\
\> \Tuple\\
\> \Expression{\tt .}\MethodName{\tt (}\Expression*{\tt )}\\
\> \UnaryOp{\tt (}\Expression{\tt )}\\
\> \BinaryOp{\tt (}\Expression{\tt ,}\Expression{\tt )}\\
\> {\tt isinstance}{\tt (}\Expression,\ClassName{\tt )}\\
\> {\tt and}{\tt (}\Expression,\Expression{\tt )} \spce \= \commentMark\ conjunction (short-circuiting)\\
\> {\tt or}{\tt (}\Expression,\Expression{\tt )}        \> \commentMark\ disjunction (short-circuiting)\\
\> {\tt each} \Iterator\ {\tt |} \Expression\\
\> {\tt some} \Iterator\ {\tt |} \Expression
\end{ctabbing}
\caption{Abstract syntax, Part 1.}
\label{fig:syntax1}
\end{figure*}

\begin{figure}[htb]
\begin{ctabbing}
\Variable\ := \= \InstanceVariable\\
\> \GlobalVariable\\
\InstanceVariable\ ::= \Expression.\Field\\
\Tuple\ ::= {\tt (}\Expression*{\tt )}\\
\Query\ := \=\Predicate \TuplePattern\\
\> \Predicate\\
\KeywordArg\ ::= \Parameter\ = \Expression\\
\UnaryOp\ ::= \=
   {\tt not}    \hspace*{3em} \= \commentMark\ Boolean negation\\
\> {\tt isTuple}           \> \commentMark\ test whether a value is a tuple\\
\> {\tt len}               \> \commentMark\ length of a tuple\\
\BinaryOp\ ::= \=
   {\tt is}           \> \commentMark\ identity-based equality\\
\> {\tt plus}         \> \commentMark\ sum\\
\> {\tt select}       \> \commentMark\ {\tt select}($t$,$i$) returns the $i$'th\twocolonly{\\ \>\> \commentMark } component of tuple $t$\\
\Pattern\ ::= \=
   \InstanceVariable\\
\> \TuplePattern\\
\TuplePattern\ ::= {\tt (}\PatternElement*{\tt )}\\
\PatternElement\ ::= \=
   \Expression\\
\> \_\\
\> \Variable\\
\> {\tt =}\Variable\\
\Iterator\ ::= \Pattern\ {\tt in} \Expression\\
\Literal\ ::= \=
  {\tt None}\\
\> \BooleanLiteral\\
\> \IntegerLiteral\\
\> ...\\
\BooleanLiteral\ ::= \=
   {\tt True}\\
\> {\tt False}\\
\IntegerLiteral\ ::= ...
\end{ctabbing}
  \caption{Abstract syntax, Part 2. Ellipses (``...'') are used for common syntactic categories whose details are unimportant.  Details of the identifiers allowed for non-terminals \ClassName, \RulesetName, \MethodName, \Parameter, \LogicVariable, \GlobalVariable, and \Field\ are also unimportant and hence unspecified, except that \ClassName\ must include {\tt set} and {\tt sequence}, and \Parameter\ must include {\tt self}.
  }
\label{fig:syntax2}
\end{figure}

\mypar{Notation in the grammar}
%
%
A symbol in the grammar is a terminal symbol if it is in typewriter font.
A symbol in the grammar is a non-terminal symbol if it is in italics.
In each production, alternatives are separated by a linebreak.
Square brackets enclose optional clauses.
{\tt *} after a non-terminal means ``0 or more occurrences''.
{\tt +} after a non-terminal means ``1 or more occurrences''.
$t\, \theta$ denotes the result of applying substitution $\theta$ to
  $t$.  We represent substitutions as (partial) functions from parameters and variables to
  expressions.

\mypar{Well-formedness requirements on programs}
In rule sets, parameters can be used only as base predicates, not derived predicates.  In rule sets defined in global scope, predicates cannot have the form {\tt self}.\Field.
In every rule in every rule set, every logic variable that appears in the conclusion must appear in a hypothesis.

Each global variable can appear as a derived predicate in at most one rule set in the program.  Each instance variable can appear as a derived predicate in at most one rule set in each class.

Invocations of methods defined using {\tt def} appear only in method call statements.  Invocations of methods defined using {\tt defun} appear only in method call expressions; we also refer to these methods as ``functions''.

\subsubsection{Constructs whose semantics is given by translation}



\mypar{Boolean operators} The Boolean operators {\tt and} and {\tt each} are eliminated as follows: {\tt $e_1$ and $e_2$} is replaced with {\tt not(not($e_1$) or not($e_2$))}, and {\tt each {\it iter} | $e$} is replaced with {\tt not(some {\it iter} | not($e$))}.


\mypar{Global variables} Global variables are replaced with instance variables by replacing each global variable $x$ with $\agv.x$, where $\agv$ is the address of an object whose fields are used to represent global variables.

\mypar{Non-variable expressions in tuple patterns} Non-variable expressions in tuple patterns are replaced with variables prefixed by ``{\tt =}''.  Specifically, for each expression $e$ in a tuple pattern that is not a variable (possibly prefixed with ``{\tt =}'') or wildcard, an assignment $v$~{\tt :=}~$e$ to a fresh variable $v$ is inserted before the statement containing the tuple pattern, and $e$ is replaced with {\tt =}$v$ in the tuple pattern.

\mypar{Wildcards} Wildcards are eliminated from tuple patterns in {\tt for} loops and comprehensions (i.e., everywhere except queries) by replacing each wildcard with a fresh variable.




\mypar{Tuple patterns in {\tt infer} statements} \hspace{0pt}{\tt infer} statements are transformed to eliminate tuple patterns in queries.  After transformation, each query is simply the name of a predicate. 
 Consider the statement {\tt 
$x_1,\ldots,x_n$ := $[e.]$infer($p_1(\pat_1),$ $\ldots,$ $p_n(pat_n), \kwargs,$ rules=$r$)}.
Let $x_{i,1},$ $\ldots,$ $x_{i,k_i}$ be the components of $\pat_i$ that are variables not prefixed by ``{\tt =}''.  Let $y_1,\ldots,y_n$ be fresh variables.  The above statement is transformed to: 
\begin{alltt}
\(y_1,\ldots,y_n\) := \([e.]\)infer(\(p_1,\ldots,p_n, \kwargs,\) rules=\(r\))
\(x_1\) := \{ (\(x_{1,1},\ldots,x_{1,k_1}\)) : \(\pat_1\) in \(y_1\) | True \}
\(\ldots\)
\(x_n\) := \{ (\(x_{n,1},\ldots,x_{n,k_n}\)) : \(\pat_n\) in \(y_n\) | True \}
\end{alltt}

\mypar{{\tt ifSome} statements} \hspace{0pt}{\tt ifSome} is statically eliminated as follows.  Consider the statement {\tt ifSome \pat\ in $e$ | $b$ : $s$}.  Let $i_1,\ldots,i_k$ be indices of elements of \pat\ that are variables not prefixed by ``{\tt =}''.  Let $x_{i_1},\ldots,x_{i_k}$ be those variables.  Let {\tt foundOne} and $x'_{i_1},\ldots,x'_{i_k}$ be fresh variables.  Let substitution $\theta$ be $[x_{i_1}:=x'_{i_1}, \ldots, x_{i_k}:=x'_{i_k}]$.  Let $\pat' = \pat\,\theta$ and $b' = b\,\theta$.
The above {\tt ifSome} statement is transformed to:
\begin{alltt}
foundOne := False
for \(pat'\) in \(e\):
  if \(b'\) and not foundOne:
    \(x\sb{i\sb{1}} := x'\sb{i\sb{1}}\)
    \(\ldots\)
    \(x\sb{i\sb{k}} := x'\sb{i\sb{k}}\)
    \(s\)
    foundOne := True
\end{alltt}

\mypar{{\tt whileSome} statements} \hspace{0pt}{\tt whileSome} is statically eliminated as follows.  Consider the statement {\tt whileSome \pat\ in $e$ | $b$ : $s$}.  Using the same definitions as in the previous item, this statement is transformed to:
\begin{alltt}
foundOne := True
while foundOne:
  foundOne := False
  for \(pat'\) in \(e\):
    if b' and not foundOne: 
      \(x\sb{i\sb{1}} := x'\sb{i\sb{1}}\)
      \(\ldots\)
      \(x\sb{i\sb{k}} := x'\sb{i\sb{k}}\)
      \(s\)
      foundOne := True
\end{alltt}


\mypar{Comprehensions}  First, comprehensions are transformed to eliminate the use of variables prefixed with ``{\tt =}''.  Specifically, for a variable {\tt x} prefixed with ``{\tt =}'' in a comprehension, replace occurrences of {\tt =x} in the comprehension with occurrences of a fresh variable {\tt y}, and add the conjunct {\tt y is x} to the Boolean condition.  Second,
all comprehensions are statically eliminated as follows.  The comprehension {\tt $x$ := \{ $e$ | $x_1$ in $e_1$, $\ldots$, $x_n$ in $e_n$ | $b$ \}}, where each $x_i$ is a pattern, is replaced with
\begin{alltt}
\(x\) := new set
for \(x\sb{1}\) in \(e\sb{1}\):
  ...
    for \(x\sb{n}\) in \(e\sb{n}\):
      if \(b\):
        \(x\).add(\(e\))
\end{alltt}

\mypar{Tuple patterns in iterators} Iterators containing tuple patterns are rewritten as iterators without tuple patterns.


Consider the existential quantification {\tt some ($e_1,
    \ldots,$ $e_n$) in $s$ | $b$}.  Let $x$ be a fresh variable.  Let
  $\theta$ be the substitution that replaces $e_i$ with {\tt
    select($x$,$i$)} for each $i$ such that $e_i$ is a variable not
  prefixed with ``{\tt =}''.  Let $\{j_1,\ldots,j_m\}$ contain the indices
  of the constants and the variables prefixed with ``{\tt =}'' in {\tt
    ($e_1, \ldots ,e_n$)}.  Let $\bar e_j$ denote $e_j$ after removing the
  ``{\tt =}'' prefix, if any.  The quantification is rewritten as {\tt some
    $x$ in $s$ | isTuple($x$) and len($x$) is $n$ and (select($x$,$j_1$),
    $\ldots$, select($x$,$j_m$)) is ($\bar e_{j_1}$, $\ldots$, $\bar
    e_{j_m}$) and $b\,\theta$}.


Consider the loop {\tt for ($e_1,\ldots,e_n$) in $e$ : $s$}.  Let $x$ and
 $S$ be fresh variables.  Let $\{i_1,\ldots,i_k\}$ contain the indices
  in {\tt ($e_1, \ldots ,e_n$)} of variables not prefixed with ``{\tt =}''.
      Let
  $\{j_1,\ldots,j_m\}$ contain the indices in {\tt ($e_1, \ldots ,e_n$)} of
  the constants and the variables prefixed with ``{\tt =}''.  Let $\bar
  e_j$ denote $e_j$ after removing the ``{\tt =}'' prefix, if any.  Note
  that $e$ may denote a set or sequence, and duplicate bindings for the
  tuple of variables $(e_{i_1},\ldots,e_{i_k})$ are filtered out if $e$ is
  a set but not if $e$ is a sequence.  The loop is rewritten as the code in
  Figure \ref{fig:elim-tuple-from-for-loop}.
 
\begin{figure}[htb]
\begin{alltt}
    \(S\) := \(e\)
    if isinstance(\(S\),set):
      \(S\) := \{ \(x\) : \(x\) in \(S\) | isTuple(\(x\)) and len(\(x\)) is \(n\)
          and (select(\(x\),\(j\sb{1}\)), \(\ldots\), select(\(x\),\(j\sb{m}\)))
             is (\(\bar{e}\sb{j\sb{1}}\), \(\ldots\), \(\bar{e}\sb{j\sb{m}}\)) \}
      for \(x\) in \(S\): 
        \(e\sb{i\sb{1}} := {\tt select}(x,i\sb{1})\)
        \(\ldots\)
        \(e\sb{i\sb{k}} := {\tt select}(x,i\sb{k})\)
        \(s\)
    else: {\rm \commentMark \(S\) is a sequence}
      for \(x\) in \(S\): 
        if (isTuple(\(x\)) and len(\(x\)) is \(n\) 
            and (select(\(x\),\(j\sb{1}\)), \(\ldots\), select(\(x\),\(j\sb{m}\)))
               is (\(\bar{e}\sb{j\sb{1}}\), \(\ldots\), \(\bar{e}\sb{j\sb{m}}\)): 
          \(e\sb{i\sb{1}} := {\tt select}(x,i\sb{1})\)
          \(\ldots\)
          \(e\sb{i\sb{k}} := {\tt select}(x,i\sb{k})\)
          \(s\)
        else: 
          skip
\end{alltt}
\caption{Translation of {\tt for} loop to eliminate tuple pattern.}
\label{fig:elim-tuple-from-for-loop}
\end{figure}

\subsection{Semantic domains}
\label{sec:domains}

The semantic domains are defined in Figure \ref{fig:domains}, using the following notation.  $D^*$ contains finite sequences of values from domain $D$.  $\Set{D}$ contains finite sets of values from domain $D$. $D_1 \pfn D_2$ contains partial functions from $D_1$ to $D_2$.  $\dom(f)$ and $\range(f)$ are the domain and range, respectively, of a partial function $f$.
  
Consider a state $\tuple{s,ht,h,\stk}$.  $s$ is the statement to be executed.  $ht$ is the heap type map; $ht(a)$ is the type of the object on the heap at address $a$.  $h$ is the heap; it maps addresses to objects.  $\stk$ is a kind of call stack: an entry is pushed on the stack when a method is called, and popped when a method returns.  However, entries on the stack do not contain bindings of method parameters to arguments (such bindings are unnecessary, because the transition rule for calls to user-defined methods substitutes the arguments into an inlined copy of the method body).  Instead, each entry contains sets of rules whose results should be automatically maintained during the method call; specifically, it contains the sets of rules defined in the class of the target object $o$ of the method call, instantiated by replacing {\tt self} with $o$.   The stack is initialized with an entry containing sets of rules defined in global scope.  That entry is never popped.  This ensures that global rule sets are always maintained.

\begin{figure}[htb]
\begin{eqnarray*}
  \Bool \negspc&=&\negspc \set{{\tt True}, {\tt False}}\\
  \Int \negspc&=&\negspc ... \\
  \Address \negspc&=&\negspc ...\\
  \Tuple \negspc&=&\negspc \Val^*\\
  \Val \negspc&=&\negspc \Bool \union \Int \union \Address \union \Tuple \union \{\none\}\\
  \Object \negspc&=&\negspc (\Field \pfn \Val) \union \Set{\Val} \union \Val^*\\
  \HeapType \negspc&=&\negspc \Address \pfn \ClassName\\
  \Heap \negspc&=&\negspc \Address \pfn \Object\\
  \State \negspc&=&\negspc \Statement \times \HeapType \times \Heap \times \Set{\Ruleset}^*
\end{eqnarray*}
\caption{Semantic domains.  Ellipses are used for semantic domains of primitive values whose details are standard or unimportant.}
  \label{fig:domains}
\end{figure}

\subsection{Extended abstract syntax}
\label{sec:extended-syntax}

Section \ref{sec:syntax} defines the abstract syntax of programs that can be written by the user.  We extend the abstract syntax to include additional forms into which programs may evolve during evaluation.  The new productions appear below.   The statement {\tt for $v$ inTuple $t$:~$s$} iterates over the elements of tuple $t$, in the obvious way.

\begin{ctabbing}
\Expression\ ::= \= {\it Address}\\
\> {\it Address}.\Field\\
\\
\Statement\ ::= {\tt for} {\it Variable} {\tt inTuple} \Tuple: \Statement
\end{ctabbing}

\subsection{Evaluation contexts}

Evaluation contexts, also called reduction contexts, are used to identify
the next part of an expression or statement to be evaluated.  An evaluation
context is an expression or statement with a hole, denoted \hole, in
place of the next sub-expression or sub-statement to be evaluated.  Evaluation contexts are defined in Figure \ref{fig:eval-context}.  Note that square brackets enclosing a clause indicate that the clause is optional; this is unrelated to the notation \hole\ for the hole.  For example, the definition of evaluation contexts for method calls implies that the expression denoting the target object is evaluated first to obtain an address (if the expression isn't already an address); then, the arguments are evaluated from left to right.  The left-to-right order holds because an argument can be evaluated only if the arguments to its left are values, as opposed to more complicated unevaluated expressions.  The definition of evaluation contexts for {\tt infer} implies that the expressions for the targets of the assignment are evaluated from left to right, then the expression for the target object, if any (i.e., if the call is for a rule set with class scope), is evaluated, and then the argument expressions are evaluated from left to right.
 


\begin{figure}[htb]
\begin{ctabbing}
\EC\ ::= \=
 \hole\\
\> (\Val*,\EC,\Expression*)\\
\> \EC.\MethodName(\Expression*)\\
\> \Address.\MethodName(\Val*,\EC,\Expression*)\\
\> \UnaryOp(\EC)\\
\> \BinaryOp(\EC,\Expression)\\
\> \BinaryOp(\Val,\EC)\\
\> {\tt isinstance}(\EC,\ClassName)\\
\> {\tt or}(\EC,\Expression)\\
\> {\tt some} \Pattern\ {\tt in} \EC\ {\tt |} \Expression\\
\> \EC.\Field\ {\tt :=} \Expression\\
\> \EC.\Field\ {\tt :=} {\tt new} \ClassName\\
\> \Address.\Field\ {\tt :=} \EC\\
\> \EC\ ; \Statement\\
\> {\tt if} \EC{\tt :} \Statement\ {\tt else:} \Statement\\
\> {\tt for} \InstanceVariable\ {\tt in} {\it \EC}{\tt :} \Statement\\
\> {\tt for} \InstanceVariable\ {\tt inTuple} \Tuple{\tt :} \EC\\
\> (\Address.\Field)*, \EC.\Field, (\Expression.\Field)* {\tt :=}\\
\> \hspace*{1em}[\Expression.]{\tt infer(}\=\Query*, \KeywordArg*,\\
\>\>{\tt rules=\RulesetName)}\\
\> (\Address.\Field)* {\tt :=} \EC.{\tt infer(}\=\Query*, \KeywordArg*,\\
\>\>{\tt rules=\RulesetName)}\\
\> (\Address.\Field)* {\tt :=}\\
\> \hspace*{1em}[\Address.]{\tt infer(}\Query*, (\Parameter=\Val)*,\\
\> \hspace*{1em}{\tt \Parameter=\EC, \KeywordArg{}*, rules=\RulesetName)}
\end{ctabbing}
   \caption{Evaluation contexts.}
   \label{fig:eval-context}
 \end{figure}

\subsection{Transition relations}
\label{sec:transition}

The transition relation for expressions has the form $ht:h \vdash e \ra e'$, where $e$ and $e'$ are expressions, $ht\in\HeapType$, and $h\in\Heap$. The transition relation for statements has the form $\sigma \ra \sigma'$ where $\sigma\in\State$ and $\sigma'\in\State$.  

Both transition relations, and some of the auxiliary functions defined below, are implicitly parameterized by the program, which is needed to look up method definitions, rule set definitions, etc.  The transition relation for expressions is defined in Figure \ref{fig:transition-expr}.  The transition relation for statements is defined in Figures \ref{fig:transition-one}--\ref{fig:transition-two}.  The context rules for expressions and statements allow the expression or statement in the evaluation context's hole to take a transition, while its context $C$ (i.e., the rest of the program) is carried along unchanged.

\mypar{Notation}
In the transition rules, $a$ matches an address, and $v$ matches a value (i.e., an element of $\Val$).

For an expression or statement $e$, $e[x := y]$ denotes $e$ with all occurrences of $x$ replaced with $y$.

A function $f$ matches the pattern $f[x \ra y]$ iff $f(x)$ equals $y$. For a function $f$, $f[x := y]$ denotes the function that is the same as $f$ except that it maps $x$ to $y$.  $\emptyfn$ denotes the empty partial function, i.e., the partial function whose domain is the empty set. For a (partial) function $f$, $f \ominus a$ denotes the function that is the same as $f$ except that it has no mapping for $a$.  For functions $f$ and $g$ with disjoint domains, $f \union g$ is their union.  For functions $f$ and $g$ with possibly overlapping domains, $f \sqcup g$ is like union but with $g$ having precedence, i.e., $(f \sqcup g)(x)=g(x)$ for $x \in \dom(f) \intersect \dom(g)$.

  
Sequences are denoted with angle brackets, e.g., $\seq{0,1,2} \in \Int^*$.  $s@t$ is the concatenation of sequences $s$ and $t$. $\first(s)$ is the first element of sequence $s$.  $\rest(s)$ is the sequence obtained by removing the first element of $s$.  $\length(s)$ is the length of sequence $s$. 

\mypar{Auxiliary functions}
$\new(c)$ returns a new instance of class $c$, for $c\in\ClassName$.
  \begin{displaymath}
    \begin{array}{@{}l@{}}
    \new(c) = \left\{\begin{array}{@{}ll@{}} 
        \emptySet & \mbox{if } c={\tt set}\\
        \emptyseq & \mbox{if } c={\tt sequence}\\
        \emptyfn & \mbox{otherwise}
      \end{array}\right.
    \end{array}
  \end{displaymath}

$\freshaddrs(h,f)$ holds if function $f$ maps its domain to distinct fresh addresses, i.e., addresses not used in heap $h$: $\freshaddrs(h,f) = \range(f) \subseteq \Address \setminus \dom(h)
\land (\forall x_1,x_2 \in \dom(f). f(x_1) \ne f(x_2))$.

$\extends(c_1,c_2)$ holds iff class $c_1$ is a descendant of class
  $c_2$ in the program's inheritance hierarchy.
 
  
$\methodDef(c, m, {\it def})$ holds iff (1) class $c$ defines
  method $m$, and {\it def} is the definition of $m$ in $c$, or (2) $c$
  does not define $m$, and {\it def} is the definition of $m$ in the
  nearest ancestor of $c$ in the inheritance hierarchy that defines $m$.

$\rulesetsg$ is the set of rule sets defined in global scope in the program.  For a rule set name $r$ in $\rulesetsg$, $\rules(r)$ is the set of rules in rule set $r$ defined in global scope.  $\rulesets(c)$ is the set of rule sets defined in class $c$ in the program.  For a rule set name $r$ in  $\rulesets(c)$, $\rules(c,r)$ is the set of rules in rule set $r$ defined in class $c$.  

For a set of rules $R$, $\basePredParams(R)$ is the set of base predicates in $R$ that are parameters.  $\basePredVars(R)$ is the set of base predicates in $R$ that are variables.  $\derivedPredVars(R)$ is the set of derived predicates in $R$ that are variables, and $\derivedPredParams(R)$ is the set of derived predicates in rules in $R$ that are parameters.  For a sequence of sets of rules $\stk$, $\derivedPredVars(\stk)$ is the set of derived predicates that are variables in any rule in $\stk$.  A variable's value is ``known'', for purpose of rule evaluation, if its value is a set, so we define $\knownVars({\it vars},ht,h) = \{ a.f \in {\it vars} \;|\; f \in \dom(h(a)) \land h(a)(f) \in \Address \land ht(h(a)(f)) = {\tt set}\}$.
For $c\in\ClassName$, $\derivedPredVars(c)$ is the set of derived predicates that are variables in any rule set defined in class $c$.  $\derivedPredVarsg$ is the set of global variables that are derived predicates in any rule set in the program.


The relation $\maintain(\theta,\thetaht,ht,h,\stk)$ defined in Figure \ref{fig:maintain} holds if, given heap type map $ht$, heap $h$, and stack $\stk$ (i.e., the last component of the state, described in Section \ref{sec:domains}), $\theta$ and $\thetaht$ are substitutions to apply to $h$ and $ht$, respectively, to express the result of automatic maintenance of the rule sets in $\stk$.  It is defined using the function $\maintsub(ht,$ $h, \stk, \newfn)$ which computes a substitution expressing the heap updates to be performed.  Its last argument $\newfn$ provides fresh addresses for sets created as a result of maintenance.  It takes $\newfn$ as an argument, instead of non-deterministically choosing fresh addresses itself, so it can be defined as a function, rather than a relation, which would be less intuitive.  To be safe, $\newfn$ specifies fresh addresses to use (if needed) for all derived predicates; no harm is done if some go unused (e.g., if some derived predicates remain undefined because they depend on base predicates with invalid values).  The function $\pi_1$ returns the first component of a tuple.

$\maintsub$ calls a function $\infsub$ to compute the result of inference for a single rule set, and uses function ${\it reduce}$ to iterate $\infsub$ over the stack $\stk$.  ${\it reduce}(f, s, v_0)$ iterates over sequence $s$, accumulating a result by recursively applying the two-argument function $f$, using $v_0$ as the initial value; it is the same as in the Python functools library.  $\infsub$'s arguments include a binding $\args$ of parameters to values (for automatic maintenance, $\args$ is the empty function). $\infsub$ returns a pair containing the substitution to apply to the heap $h$ and a function $\result$ that maps each defined derived predicate (i.e., each derived predicate that depends only on base predicates whose values are sets) in the rule set to its value.   $\infsub$ uses the following auxiliary functions.  $\slice(R, \knownbps)$ returns a subset of the given set of rules $R$ obtained by first identifying the set of derived predicates that depend only on base predicates in $\knownbps$, and then returning only the rules on which those derived predicates depend. $\evalrules(R)$ evaluates the given set of rules $R$ and returns a function from the set of predicates that appear in the rules to their meanings, represented as sets of tuples. $\updatevar(ht,h,\newfn,$ $a.f,S)$, defined in Figure \ref{fig:maintain}, returns a substitution that updates variable $a.f$ to refer to a set with content $S$.  If $a.f$ already contains an address, the object at that address is changed to be a set with content $S$, otherwise $a.f$ is assigned a fresh address which is updated to contain a set with content $S$.

$\legalassign(ht,a,f)$ holds if assigning to field $f$ of the object with address $a$ is legal, in the sense that $a$ refers to an object with fields (not an instance of a built-in class without fields), and $a.f$ is not a derived predicate of any rule set.  
$\legalassign(ht,a,f) = ht(a) \not\in\set{{\tt set}, {\tt sequence}} \land
((a = \agv \land \agv.f \not\in\derivedPredVarsg \lor 
(a \ne \agv \land {\tt self.f} \not\in\derivedPredVars(ht(a)))$.

\mypar{Notes}
Transition rules for methods of the pre-defined classes {\tt set} and {\tt sequence} are similar in style, so only one representative example is given, for {\tt set.add}.  Note that $\maintain$ needs to be called only in transition rules for methods of {\tt set} that update the content of the set. 

Informally, the transition rule for invoking a method in a user-defined class inlines a copy of the method body $s$ that has been instantiated by substituting argument values for parameters, appends a {\tt return} statement to the method body to mark the end of the method call, updates the stack $\stk$ by appending instantiated copies of the rule sets defined in class $ht(a)$ (i.e., the class of the target object of the method call), uses the auxiliary relation $\maintain$ to identify substitutions $\thetaht$ and $\theta$ that reflect the effect of automatic maintenance of rule sets on the heap type map and heap, and applies those substitutions to obtain the updated heap type map and heap.

The transition rule for an explicit call to {\tt infer} on a rule set with class scope instantiates the named rule set $r$ using the given values for the rule set's parameters, uses auxiliary function $\infsub$ to evaluate the instantiated rule set, and uses $\maintain$ to determine the effects of automatic maintenance.  To understand the rule, note that $\newfn$ provides fresh addresses for sets created as a result of this call to {\tt infer}; ${\it args}$ captures the values of keyword arguments; $R$ is the set of rules in $r$ with {\tt self} instantiated with the target object $a$; $\theta$ is a substitution to apply to the heap to update the values of derived predicates of $r$ that are variables; $\result$ maps each defined derived predicate of $R$ (i.e., each derived predicate of $R$ that depends only on base predicates whose values are sets) to its value; {\it definedParams} and {\it undefParams} are the sets of derived predicates of $r$ that are parameters and whose values are defined and undefined (i.e., \none), respectively, as a result of this call to {\tt infer}; $\theta_{\it Q}$ and $\theta_{\it Qundef}$ are substitutions to apply to the heap to update the values of $a_1.f_1,\ldots,a_n.f_n$; and $\thetaht$ is a substitution to apply to the heap type map $ht$ to update the types of heap locations containing sets created by this call to {\tt infer}.


A rule set that defines a global variable as a derived predicate that depends on a predicate of the form {\tt self}.\Field{} has two notable effects.  First, when there are nested calls on different target objects that are instances of a class $c$ that defines such a rule set, the substitutions generated by $\infsub$ from the  stack entries for those nested calls have overlapping domains; this is why $\maintsub$ combines them using $\sqcup$ instead of $\union$.  In particular, they are combined so that the innermost call (topmost stack frame) takes precedence.  Second, when the stack does not contain any call to a method of $c$, the global variable is set to \none.  Thus, returning from a method call can have the side-effect of changing the value of a global variable to \none.

Automatic maintenance has the effect of evaluating all rule sets using values of their base predicates in the current state, and then updating their derived predicates in the next state, like a single ``parallel assignment'' statement, even if there are dependencies between rule sets.  For example, suppose there are rule sets $r_1$ and $r_2$ in global scope, and some derived predicate $x$ of $r_1$ is also a base predicate of $r_2$.  In a transition from state $\sigma_1$ to state $\sigma_2$, $r_2$ is evaluated using the value of $x$ in $\sigma_1$, not its value in $\sigma_2$.




\begin{figure}[htb]
\begin{displaymath}
\begin{array}{@{}l@{}}
\updatevar(ht,h,\newfn,a.f,S) = \\
\sci \mbox{if } h(a)(f) \in \Address \mbox{ then } [h(a)(f) := S] \\
\sci \mbox{else } [a := h(a)[f := \newfn(a.f)], \newfn(a.f) := S]\\
\\
\infsub(ht, h, \newfn, \args, R) = \\
\sci \begin{array}{@{}l@{}}
\mbox{let } {\it vars} = \knownVars(\basePredVars(rules),ht,h)\\
\mbox{and } {\it params} =  \set{X : X \in \dom(\args) \land ht(\args(X))={\tt set}}\\
\mbox{and } {\it R'} = \slice(R, {\it vars} \union {\it params})\\
\mbox{and } {\it facts}_V = \set{a.f(v) : a.f \in {\it vars} \land v \in h(a)(f)}\\
\mbox{and } {\it facts}_P = \set{X(v) : X \in {\it params} \land v \in h(\args(X))}\\
\mbox{and } \result = \evalrules(R' \union {\it facts}_V \union {\it facts}_P)\\
\mbox{and } {\it definedDPs} = \derivedPredVars(R)\intersect\dom(\result)\\
\mbox{and } {\it undefDPs} = \derivedPredVars(R)\setminus\dom(\result)\\
\mbox{and } \theta = \union_{v\in {\it definedDPs}} \updatevar(ht,h,\newfn,v,\result(a.f))\\
\mbox{and } \theta_{\it undef} = \union_{v\in {\it undefDPs}} [a := h(a)[f := \none]]\\
\mbox{in } \tuple{\theta \union \theta_{\it undef}, \result}
\end{array}\\
\\
\maintsub(ht, h, \newfn, \stk) = \\
\sci \mbox{let } f(\theta, RS) = \theta \sqcup (\union_{R \in RS} \pi_1(\infsub(ht,h,\newfn, f_0, R)))\\
\sci \mbox{in } {\it reduce}(f, \stk, f_0)\\
\\
\maintain(\theta,\thetaht,ht,h,\stk) = \\
\sci \begin{array}{@{}l@{}}
\exists \newfn \in \derivedPredVars(\stk) \pfn \Address .\\
\freshaddrs(ht,\newfn) \land
\theta = \maintsub(ht,h,\stk,\newfn)\\
{} \land \thetaht = [a := {\tt set} : a \in \dom(\theta) \land \theta(a) \in \Val^*]
\end{array}
\end{array}
\end{displaymath}
\caption{Definition of $\maintain$ and related functions.}
\label{fig:maintain}
\end{figure}

\begin{figure}[tbp]
\begin{displaymath}
\begin{array}{@{}l@{}}
\commentS{field access}\\
ht:h \vdash a.f \ra h(a)(f) \spce \IF f \in \dom(h(a))\\
\\
\commentS{invoke function in user-defined class}\\
ht:h \vdash a{\tt .}m(v_1,\ldots,v_n) \ra e[{\tt self}:=a, x_1:=v_1, \ldots, x_n:=v_n]\\
\sci \IF \methodDef(ht(a), m, {\tt defun}~m(x_1,\ldots,x_n)~e)\\
\\
\commentS{invoke function in pre-defined class (example)}\\
ht:h \vdash a{\tt .any()} \ra v
\spce \IF ht(a)={\tt set} \land v \in h(a)\\
ht:h \vdash a{\tt .any()} \ra {\tt None}
\spce \IF ht(a)={\tt set} \land h(a)=\emptyset\\
\\
\commentS{unary operations}\\
ht:h \vdash {\tt not(True)} \ra {\tt False}\\
ht:h \vdash {\tt not(False)} \ra {\tt True}\\
ht:h \vdash {\tt isTuple(}v{\tt )} \ra {\tt True} \spce 
\mbox{if $v$ is a tuple}\\
ht:h \vdash {\tt isTuple(}v{\tt )} \ra {\tt False} \spce 
\mbox{if $v$ is not a tuple}\\
ht:h \vdash {\tt len(}v{\tt )} \ra n  \spce 
\mbox{if $v$ is a tuple with $n$ components}\\
\\
\commentS{binary operations}\\
ht:h \vdash {\tt is(}v_1,v_2{\tt )} \ra {\tt True} \\
\sci \mbox{if $v_1$ and $v_2$ are the same (identical) value}\\
\\
ht:h \vdash {\tt plus(}v_1,v_2{\tt )} \ra v_3 \\
\sci \IF v_1\in\Int \land v_2\in\Int \land v_3 = v_1 + v_2\\
\\
ht:h \vdash {\tt select(}v_1,v_2{\tt )} \ra v_3\\
\sci \IF v_2\in\Int \land v_2>0 \land
\mbox{($v_1$ is a tuple with length at least $v_2$)}\\
\sci {} \land \mbox{($v_3$ is the $v_2$'th component of $v_1$)}\\
\\
\commentS{isinstance}\\
ht:h \vdash {\tt isinstance(}a, c{\tt )} \ra {\tt True} \spce \IF ht(a)=c\\
ht:h \vdash {\tt isinstance(}a, c{\tt )} \ra {\tt False} \spce \IF ht(a)\ne c\\
\\
\commentS{disjunction}\\
ht:h \vdash {\tt or(True}, e{\tt )} \ra {\tt True}\\
ht:h \vdash {\tt or(False}, e{\tt )} \ra e\\
\\
\commentS{existential quantification}\\
ht:h \vdash {\tt some}~x~{\tt in}~a~|~e ~~\ra~~ e[x:=v_1]~{\tt or }~\cdots~{\tt or}~e[x:=v_n]\\
\sci \IF (ht(a)={\tt sequence} \land h(a)=\seq{v_1,\ldots,v_n}) \\
\sci {} \lor (ht(a)={\tt set} \land \seq{v_1,\ldots,v_n}~\mbox{is a linearization of}~h(a))
\end{array}
\end{displaymath}
  \caption{Transition relation for expressions.}
  \label{fig:transition-expr}
\end{figure}

\begin{figure}[tbp]
\setstretch{0.96}
\begin{displaymath}
\begin{array}{@{}l@{}}
\commentS{context rule for expressions}\\
\dfrac{h(a):ht \vdash e \ra e'}{\tuple{C[e], ht, h, ch, mq} \ra \tuple{C[e'], ht, h, ch, mq}}\\
\\
\commentS{context rule for statements}\\
\dfrac{\tuple{s, ht, h, ch, mq} \ra \tuple{s', ht', h', ch', mq'}}{\tuple{C[s], ht, h, ch, mq} \ra \tuple{C[s'], ht', h', ch', mq'}}\\
\\
\commentS{field assignment}\\
\tuple{a{\tt .}f \;{\tt :=}\; v, ht, h[a \ra o], \stk}\\
{} \ra \tuple{{\tt skip}, ht\,\thetaht, h'\,\theta, \stk}\\
\sci{} \IF \legalassign(ht,a,f) \land h' = h[a := o[f := v]]\\
\sci {} \land \maintain(\theta,\thetaht,ht,h',\stk)\\
\\
\commentS{object creation}\\
\tuple{a{\tt .}f \;{\tt :=}\; {\tt new}~c, ht, h[a \ra o], \stk}\\
{}\ra \tuple{{\tt skip}, ht'\thetaht, h'\,\theta, \stk}\\
\sci \IF a'\not\in\dom(ht) \land a' \in \Address \land \legalassign(ht,a,f)\\
\sci {} \land ht' = ht\,[a' := c] \land h' =  h[a := o[f := a'], a' := \new(c)]\\
\sci {} \land \maintain(\theta,\thetaht,ht',h',\stk)
\\
\\
\commentS{sequential composition}\\
\tuple{{\tt skip;}\, s, ht, h, \stk} \ra
\tuple{s, ht, h, \stk}\\
\\
\commentS{conditional statement}\\
\tuple{{\tt if}~{\tt True:}~s_1~{\tt else:}~s_2, ht, h, \stk} \ra
\tuple{s_1, ht, h, \stk}\\
\\
\tuple{{\tt if}~{\tt False:}~s_1~{\tt else:}~s_2, ht, h, \stk} \ra
\tuple{s_2, ht, h, \stk}\\
\\
\commentS{for loop}\\
\tuple{{\tt for}~x~{\tt in}~a{\tt :}~s, ht, h, \stk}\\
{} \ra
\tuple{{\tt for}~x~{\tt inTuple}~{\tt (}v_1,\ldots,v_n{\tt ):}~s, ht, h, \stk}\\
\sci \IF (ht(a)={\tt sequence} \land h(a)=\seq{v_1,\ldots,v_n})\\
\sci {} \lor
    (ht(a)={\tt set} \land \seq{v_1,\ldots,v_n}~\mbox{is a linearization of}~h(a))\\
\\
\tuple{{\tt for}~x~{\tt inTuple}~{\tt (}v_1,\ldots,v_n{\tt ):}~s, ht, h, \stk}\\
{}\ra \tuple{s[x:=v_1]; {\tt for}~x~{\tt inTuple}~{\tt (}v_2,\ldots,v_n{\tt ):}~s, ht, h, \stk}\\
\\
\tuple{{\tt for}~x~{\tt inTuple}~{\tt ():}~s, ht, h, \stk} \ra
\tuple{{\tt skip}, ht, h, \stk}\\
\\
\commentS{while loop}\\
\tuple{{\tt while}~e{\tt :}~s, ht, h, \stk} \\
{} \ra
\tuple{{\tt if}~e{\tt :}~{\tt (}s{\tt ;}~{\tt while}~e{\tt :}~s{\tt )}~{\tt else:}~{\tt skip}, ht, h, \stk}
\end{array}
\end{displaymath}
  \caption{Transition relation for statements, Part 1.}
  \label{fig:transition-one}
\end{figure}

\begin{figure}[tbp]
\begin{displaymath}
\begin{array}{@{}l@{}}
\commentS{invoke method in pre-defined class (example)}\\
\tuple{a{\tt .add(}v_1{\tt )}, ht, h, \stk}\ra
\tuple{{\tt skip}, ht\thetaht, h'\theta, \stk}\\
\sci \IF ht(a)={\tt set} \land h' = h[a := h(a)\union\set{v_1}]\\
\sci {} \land \maintain(\theta,\thetaht,ht,h',\stk)\\
\\
\commentS{invoke method in user-defined class}\\
\tuple{a{\tt .}m(v_1,\ldots,v_n), ht, h, \stk}\\
{} \ra
\tuple{s[{\tt self}:=a, x_1:=v_1, \ldots, x_n:=v_n]{\tt ;}\, {\tt return}, ht\,\thetaht, h\,\theta, \stk'}\\
\sci\IF ht(a)\not\in\set{{\tt set}, {\tt sequence}}\\
\sci {} \land \methodDef(ht(a), m, {\tt def}~m{\tt (}x_1,\ldots,x_n{\tt )}~s)\\
\sci {} \land \stk' = \stk@\seq{\set{\rules(ht(a),r)[{\tt self}:=a] : r \in \rulesets(ht(a))}}\\
\sci {} \land \maintain(\theta,\thetaht,ht,h,\stk')\\
\\
\commentS{return from call to method in user-defined class}\\
\tuple{{\tt return}, ht, h, \seq{s_1,\ldots,s_n}} \ra 
\tuple{{\tt skip}, ht\,\thetaht, h\,\theta, \stk'}\\
\sci \IF \stk' = \seq{s_1,\ldots,s_{n-1}} \land \maintain(\theta,\thetaht,ht,h,\stk')\\
\\
\commentS{invoke {\tt infer} on a rule set defined in class scope}\\
\ltuple a_1.f_1,\ldots,a_n.f_n := a{\tt .infer(}q_1,\ldots,q_n, x_1=v_1, \ldots, x_k=v_k,\\
\tpli {\tt rules=}r{\tt )}, ht, h, \stk\rtuple
\ra \tuple{{\tt skip}, ht\, \thetaht\thetaht', h\, \theta \theta_Q\theta_{\it Qundef}\theta', \stk}\\
\sci\IF r \in \rulesets(ht(a))\\
\sci 
\begin{array}{@{}l@{}}
{} \land (\forall i \in \set{1..n}. \legalassign(ht,a_i,f_i))\\
{} \land
\newfn \in (\derivedPredVars(\stk) \union \set{q_1,\ldots,q_n}) \pfn \Address\\
{} \land \freshaddrs(ht,\newfn)\\
{} \land \args = f_0[x_i := v_i : i \in \set{1..k}]\\
{} \land R = \rules(ht(a),r)[{\tt self}:=a]\\
{} \land \tuple{\theta,\result} = \infsub(ht, h, \newfn, \args, R)\\
{} \land {\it definedParams} = \derivedPredParams(R) \intersect \dom(\result)\\
{} \land {\it undefParams} = \derivedPredParams(R) \setminus \dom(\result)\\
{} \land \theta_{\it Q} = \UNION_{q_i \in {\it definedParams}}
\begin{array}[t]{@{}l@{}}
 [a_i := h(a_i)[f_i := \newfn(q_i)],\\
 \newfn(q_i) := \result(q_i)]
\end{array}\\
{} \land \theta_{\it Qundef} = 
\UNION_{q_i \in {\it undefParams}} [a_i := h(a_i)[f_i := \none]]\\
{} \land \thetaht = [a := {\tt set} : 
\begin{array}[t]{@{}l@{}}
(a \in \dom(\theta) \land \theta(a) \in \Val^*)\\
{} \lor (a \in \dom(\theta_Q) \land \theta_Q(a) \in \Val^*]
\end{array}\\
{} \land \maintain(\theta',\thetaht',ht\,\thetaht, h\,\theta\theta_Q\theta_{\it Qundef},\stk)
\end{array}\\
\\
\commentS{invoke {\tt infer} on a rule set defined in global scope}\\
\mbox{same as previous rule, except replace $a{\tt .infer}$ with {\tt infer},}\\
\mbox{replace conjunct $r \in \rulesets(ht(a))$ with $r \in \rulesetsg$,}\\
\mbox{replace $\rules(ht(a),r)$ with $\rules(r)$,}\\
\mbox{and omit substitution $[{\tt self}:=a]$.}
\end{array}
\end{displaymath}
  \caption{Transition relation for statements, Part 2.}
  \label{fig:transition-two}
\end{figure}


\mypar{Executions}
An execution is a sequence of transitions $\sigma_0 \ra \sigma_1 \ra
\sigma_2 \ra \cdots$ such that $\sigma_0$ is the initial state of the program, given by $\sigma_0 = ( s_0, f_0, f_0, \seq{\{ \rules(r) \;|\; r \in \rulesetsg \} } )$, where $s_0$ is the top-level statement in the program.

Informally, execution of a program may eventually (1) terminate (i.e., the statement in the first component of the state becomes {\tt skip}, indicating that there is nothing left for the process to do), (2) get stuck (i.e., the statement is not {\tt skip}, and the process has no enabled transitions, due to an error), or (3) run forever due to an infinite loop or infinite recursion.  Examples of errors that cause a program to get stuck are trying to select a component from a value that is not a tuple, and trying to access a non-existent field of an object.  For brevity and simplicity, we designed the semantics so that errors simply halt execution; the semantics could easily be extended to indicate exactly what error occurred.




\oops{}

\end{document}